%% file: NCFT_Draft_v5.tex
\documentclass[11pt,a4paper]{article}
\pdfoutput=1
\usepackage{jheppub}
\usepackage{booktabs}
\usepackage[english]{babel}
\usepackage{amsmath,amssymb,amsbsy,amstext, amsthm, simplewick}
\usepackage{hyperref}
\usepackage{graphicx}
\usepackage[export]{adjustbox}
\usepackage{amsfonts}
\usepackage{amssymb}
\usepackage[small]{caption}
\usepackage{upgreek}
\usepackage{cancel}
\usepackage{color}
\usepackage{slashed}
\usepackage[dvipsnames]{xcolor}
\usepackage{subcaption}
\usepackage{tikz-cd}
\usepackage{color}
\usepackage{soul}
\usepackage{dsfont}
\usepackage[force]{feynmp-auto}
\usepackage[utf8]{inputenc}

\newcommand{\half}{\frac{1}{2}}

\usepackage{letltxmacro}
\makeatletter
\let\oldr@@t\r@@t
\def\r@@t#1#2{%
	\setbox0=\hbox{$\oldr@@t#1{#2\,}$}\dimen0=\ht0
	\advance\dimen0-0.2\ht0
	\setbox2=\hbox{\vrule height\ht0 depth -\dimen0}%
	{\box0\lower0.4pt\box2}}
\LetLtxMacro{\oldsqrt}{\sqrt}
\renewcommand*{\sqrt}[2][\ ]{\oldsqrt[#1]{#2}}
\makeatother
\usepackage{empheq}
\newlength\dlf  

\usepackage{subfiles}

\title{IR Dynamics from UV Divergences: \\ UV/IR Mixing, NCFT, and the Hierarchy Problem}
\author[]{Nathaniel Craig}
\author[]{\& Seth Koren}

\affiliation[]{Department of Physics, University of California, Santa Barbara, 
CA 93106, USA}

\abstract{The persistence of the hierarchy problem points to a violation of effective field theory expectations. A compelling possibility is that this results from a physical breakdown of EFT, which may arise from correlations between ultraviolet (UV) and infrared (IR) physics. To this end, we study noncommutative field theory (NCFT) as a toy model of UV/IR mixing which generates an emergent infrared scale from ultraviolet dynamics. We explore the range of such theories where ultraviolet divergences are transmogrified into infrared scales, focusing particularly on the properties of Yukawa theory, where we identify a new infrared pole accessible in the $s$-channel of the Lorentzian theory. We further investigate the interplay between UV-finiteness and UV/IR mixing by studying properties of the softly-broken noncommutative Wess-Zumino model as soft terms are varied relative to the cutoff.  
While the Lorentz violation inherent to noncommutative theories may limit their direct application to the hierarchy problem, these toy models provide general lessons to guide the realization of UV/IR mixing in more realistic theories.
}

\begin{document}

\maketitle

\section{Introduction}\label{sec:intro}
\subfile{intro_v5}

\section{Noncommutative Field Theory}\label{sec:review}
\subfile{review_v5}

\section{Real Scalar \texorpdfstring{$\phi^4$}{Phi4} Theory}\label{sec:phi4}
\subfile{phi4_v5}

\section{Yukawa Theory}\label{sec:yukawa}
\subfile{yukawa_v5}

\section{Softly-broken Wess-Zumino Model}\label{sec:wesszumino}
\subfile{wesszumino_v5}

\section{Whence UV/IR Mixing?}\label{sec:lessons}
\subfile{lessons_v5}

\section{Conclusions} \label{sec:conclusions}
\subfile{conclusions_v5}

\section*{Acknowledgements}
We thank Nima Arkani-Hamed, Matthew Brown, Andy Cohen, Tim Cohen, Brianna Grado-White, Alex Kinsella, Harold Steinacker, Terry Tomboulis, Timothy Trott, and Yue Zhao for valuable discussions. We are grateful to Tim Cohen, Isabel Garcia-Garcia, and Robert McGehee for comments on a draft of this manuscript. This work is supported in part by the US Department of Energy under the Early Career Award DE-SC0014129 and the Cottrell Scholar Program through the Research Corporation for Science Advancement.

\appendix
\section{Wilsonian Interpretations from Auxiliary Fields}\label{app:auxfield}
\subfile{auxfield_v5}


\bibliography{ncft}
\bibliographystyle{JHEP}

\end{document}

%% file: intro_v5.tex
At its heart, the electroweak hierarchy problem is a question of how an infrared (IR) scale can emerge from an ultraviolet (UV) scale without fine-tuning of UV parameters. Given the sensitivity of the Standard Model Higgs mass to UV scales, the expectation of effective field theory (EFT) is that the two should coincide. Conventional solutions to the hierarchy problem introduce both symmetries that control UV contributions to the Higgs potential and dynamics that generate IR contributions, leading to considerable structure at the weak scale and correspondingly sharp experimental tests. Ongoing exploration of the weak scale has given no evidence for these solutions, despite their theoretical soundness. 

In the face of increasingly powerful LHC data in excellent agreement with the Standard Model, it's worth taking seriously the possibility that Nature may be leading us to the conclusion that \textit{there is no new physics at the weak scale}. While this is often taken to suggest the existence of considerable fine-tuning in the Higgs potential, here we pursue an alternative idea. Perhaps the apparent violation of EFT expectations at the weak scale is a sign of the breakdown of EFT itself.

This statement is not as radical as it may at first seem. That EFT must eventually break down is not a new idea; it has long been known that gravity contains low-energy effects which cannot be understood in the context of EFT. The fact that black holes radiate at temperatures inversely proportional to their masses \cite{Hawking:1974sw} necessitates some sort of `UV/IR mixing' in gravity --- infrared physics must somehow `know about' heavy mass scales in violation of a na\"{i}ve application of decoupling. As a perhaps-more-fundamental \textit{raison d'\^{e}tre} for such behavior, the demand that observables in a theory of quantum gravity must be gauge-(that is, diffeomorphism-)invariant dictates that they must be nonlocal (see e.g. \cite{Torre:1993fq,Giddings:2005id,Donnelly:2015hta,Donnelly:2016rvo,Giddings:2018umg}), again a feature which standard EFT techniques do not encapsulate. In view of this, the conventional position is that EFT should remain a valid strategy up to the Planck scale, at which quantum gravitational effects become important. But once locality and decoupling have been given up, how and why are violations of EFT expectations to be sequestered to inaccessible energies? Indeed, the `firewall' argument \cite{Almheiri:2012rt} evinces tension with EFT expectations in semiclassical gravity around black hole backgrounds at arbitrarily low energies and curvatures.

That quantum gravitational effects will affect infrared particle physics is likewise not a new idea. This has been the core message of the Swampland program \cite{Vafa:2005ui}, which has been cataloging --- to varying degrees of concreteness and certainty --- ways in which otherwise allowable EFTs may conjecturally be ruled out by quantum gravitational considerations. These are EFTs which would look perfectly sensible and consistent to an infrared effective field theorist, yet the demand that they be UV-completed to theories which include Einstein gravity reveals a secret inconsistency. While this is powerful information, the extent to which the UV here meddles with the IR is relatively minor --- just dictating where one must live in the space of infrared theories. Even so, they have been found to have possible applications to SM puzzles, including the hierarchy problem \cite{Cheung:2014vva, Ooguri:2016pdq, Ibanez:2017kvh, Ibanez:2017oqr, Hamada:2017yji, Lust:2017wrl, Gonzalo:2018tpb, Gonzalo:2018dxi, Craig:2018yvw, Craig:2019fdy}. 

In theory far more flagrant violations of low-energy expectations are permissible --- that is, the extent to which quantum gravitational violation of EFT will affect the infrared of our universe is not at all certain. Of course any proposal to see new effects from a breakdown of EFT must contend with the rampant success of the SM EFT in the IR. Certainly a violation of EFT must both come with good reason and be deftly organized to spoil only those observed EFT puzzles. For the former, the need for quantum gravity is obviously compelling. As to the latter, it is interesting to note that the most pressing mysteries involve the relevant parameters in the SM Lagrangian.

Ultimately, our ability to address the hierarchy problem through quantum gravitational violations of EFT is limited by our incomplete understanding of quantum gravity. This motivates finding non-gravitational toy models that violate EFT expectations on their own, providing a calculable playground in which to better understand the potential consequences of UV/IR mixing. In this work we pursue the idea that UV/IR mixing may have more direct effects on the SM by considering noncommutative field theory (NCFT) as such a toy model. These theories model physics on spaces where translations do not commute \cite{Snyder:1946qz,Connes:1994yd}, and have many features amenable to a quantum gravitational interpretation --- indeed, noncommutative geometries have been found arising in various limits of string theory \cite{Connes:1997cr,Douglas:1997fm,Seiberg:1999vs,Myers:1999ps}.\footnote{Noncommutative branes arising in gauge theory matrix models have also been found to contain emergent gravitational effects, and so have been suggested as novel quantum theories of gravity \cite{Rivelles:2002ez,Yang:2004vd,Yang:2006hj,Yang:2006mn,Steinacker:2007dq,Steinacker:2008ri,Steinacker:2008ya,Grosse:2008xr,Klammer:2008df,Steinacker:2009mp}. We do not pursue this perspective here, but refer the reader to \cite{Steinacker:2010rh} for a review of this approach.} 

This noncommutativity bears out the general expectation that the general-relativistic notion of spacetime should break down in a theory of quantum gravity \cite{DeWitt:1962cg}. Its realization here leads directly both to UV/IR mixing in the form of a violation of decoupling and to nonlocal effects in interactions. This gives rise to many interesting effects, but particularly fascinating for our purposes is that UV divergences present in the S-matrix elements of QFTs on commutative spaces can be transmogrified into \textit{new infrared poles} in the corresponding field theory on noncommutative space \cite{Minwalla:1999px}. An effective field theorist living in a noncommutative space would have no way to understand the appearance of this infrared scale; its existence is intrinsically linked to the geometry of spacetime and to the far UV of the theory. Such an effective field theorist would see a surprising lack of new physics accompanying this pole to explain its presence. 

It is clear from the outset that the direct application of NCFT to understand the hierarchy problem is immediately hindered by the Lorentz invariance violation which is inherent to these theories. Precisely how fatal this might be is not entirely clear; results regarding the extent to which `generic' Lorentz violation is empirically ruled out \cite{Collins:2004bp} are partly circumvented here by the fact that the Lorentz violation is \textit{not} generic, but comes as part of some larger structure. In this case the novel effects of UV/IR mixing in fact only appear in nonplanar loop diagrams \cite{Filk:1996dm} and care is required when interpreting EFT constraints on Lorentz violation --- a point we will emphasize in Section \ref{sec:review}. Even so, it is difficult to imagine that observed properties of the weak scale and the wide range of constraints on Lorentz violation leave room for NCFT to be directly relevant to puzzles of the Standard Model.

Thus we make no claim about having solved the hierarchy problem. The value of this work is in the exploration of this toy model of UV/IR mixing, which possesses the intriguing feature that ultraviolet dynamics generate a scale whose lightness would be baffling to an effective field theorist. As this is the only model (of which we are aware) with this feature --- and this feature, at the level of words, increasingly matches the experimental situation with the Higgs --- it's worth understanding its appearance in as much detail as possible.

To make this work self-contained for the contemporary particle theorist, we begin with an  extensive introduction. In Section \ref{sec:review}, we review quantum field theory on noncommutative spaces with an emphasis on the violation of EFT expectations. In Section \ref{sec:phi4} we use this technology to go over the classic result of \cite{Minwalla:1999px} which first identified this emergent infrared pole in a Euclidean $\phi^4$ theory. We compute also the effect in dimensional regularization to evince the regularization-independence of the UV/IR mixing effects.

In Section \ref{sec:yukawa} we ask how general the effect of UV/IR mixing is within NCFT, which leads us to study noncommutative Yukawa theory in detail. We find that the scalar propagator again develops a new infrared pole at one loop, in contrast with previous work. Intriguingly, the pole in this case is accessible in $s$-channel scattering in the Lorentzian theory, making Yukawa theory a promising setting for probing phenomenological consequences of UV/IR mixing.

In Section \ref{sec:wesszumino} we upgrade our model to the softly-broken Wess-Zumino model to study the interplay between UV-finiteness and UV/IR mixing effects. When the fermion is kept in the spectrum of the theory below the cutoff, the lack of UV sensitivity of the field theory removes the light pole. As the fermion is taken above the cutoff, an effective theorist again sees effects past those observed in Wilsonian EFT. These results are expected, but this model affords us a concrete demonstration that UV/IR mixing can only have interesting low-energy effects if the field theory is UV sensitive, and puts this naturalness strategy in stark contrast to conventional approaches. Of course, this also makes addressing the hierarchy problem with UV/IR mixing a potentially Pyrrhic victory: to generate an IR scale, the field theory alone cannot be fully predictive.

Finally, in Section \ref{sec:lessons} we examine the appearance of the emergent light pole in NCFT from more general arguments, so as to ascertain the relative importance of nonlocality and Lorentz-violation for these effects. The conclusion is inevitably that in this case the two are inexorably linked, and no strong conclusion about the possibility of finding a light pole in a theory with only one or the other is available. However, we provide some direction toward future explorations into both of these possibilities. We wrap up in Section \ref{sec:conclusions}.

%% file: review_v5.tex
In this section we review the salient features of the formulation of noncommutative field theories and the standard formalism for studying their perturbative physics. Useful general references for this background include \cite{Szabo:2001kg,Douglas:2001ba}. Readers familiar with NCFT may wish to skip to Section \ref{sec:phi4}, but we emphasize that our interest is necessarily non-perturbative in the parameter controlling the noncommutativity, unlike much of the earlier phenomenological literature.

Physics on noncommutative spaces involves the introduction of a nonzero commutator between position operators
\begin{equation}\label{eqn:ncdef}
\left[\hat{x}_\mu,\hat{x}_\nu\right] = i \theta_{\mu\nu},
\end{equation}

\noindent where we will refer to $\theta_{\mu\nu} = - \theta_{\nu\mu}$ as the noncommutativity tensor, and we emphasize that it is covariant under Lorentz transformations. So while it does break Lorentz invariance, it only does so in the way that turning on a magnetic field in your lab chooses a preferred frame. This basic definition is reminiscent of the introduction of a nonzero commutator in passing from classical mechanics to quantum mechanics. Indeed much of the structure is precisely analogous, including importantly the construction of noncommutative versions of familiar commutative theories via a quantization map. At an even more basic level, the above nonzero commutator induces an uncertainty relation
\begin{equation}
\Delta \hat{x}_\mu \Delta \hat{x}_\nu \geq \frac{\left|\theta_{\mu\nu}\right|}{2},
\end{equation}

\noindent which immediately makes apparent the presence of UV/IR mixing in this theory. If you attempt to create a wavepacket which is very small in one direction it will necessarily be elongated in another, and so we see already the non-trivial mixing of UV and IR modes. This clearly violates the separation of scales which is baked in to EFT. Thus purely from the defining relation of noncommutative geometry, we see already an indication that noncommutative theories should violate EFT expectations. 

Field theories on this space may be conveniently formulated in terms of fields that are functions of {\it commuting} coordinates imbued with a new field product, known as a Groenewold-Moyal product (or star-product), with position-space representation

\begin{equation}\label{eqn:starprod}
f(x) \star g(x) = \left. \exp\left(\frac{i}{2} \theta_{\mu\nu} \partial_y^\mu \partial_z^\nu\right) f(y) g(z) \right|_{y = z = x} = f(x) \exp\left(\frac{i}{2} \overleftarrow{\partial}^\mu \theta_{\mu\nu} \overrightarrow{\partial}^\nu\right) g(x).
\end{equation}

It is important to observe that this is a nonlocal product, since it contains an infinite series of derivative operators. So we see again that one of the tenets of EFT has been violated by our basic definition of field theory on noncommutative spaces.

With this in hand we may now write down noncommutative versions of familiar theories \textit{in terms of commuting coordinates}, which will then allow us to use normal QFT methods to analyze them. First note that this noncommutative quantization will not affect the quadratic part of the tree-level action due to momentum conservation and the antisymmetry of the noncommutativity tensor. For the interacting part of the action the effects of noncommutative quantization are not so trivial, but are easy to analyze classically. As an example, for a simple $\phi^n$ theory we find
\begin{equation}
\mathcal{L}^{(NC)}_\text{int} = \frac{\lambda}{n!} \overbrace{\phi(x)\star\phi(x)\star\cdots\star\phi(x)}^{\text{n copies}}.
\end{equation}

Note, importantly, that the star-product has endowed our vertices with a notion of ordering, as it is only cyclically invariant. If we now Fourier transform the action to momentum space, we find that we can account for the effects of quantization on the tree-level action with a simple modification of the momentum-space vertex factor:
\begin{equation}
\tilde{V}\left(k_1,\dots,k_n\right) = \delta\left(k_1 + \dots + k_n\right)\exp\left(\frac{i}{2} \sum\limits_{i<j}^{n}k_i^\mu k_j^\nu \theta_{\mu\nu}\right).
\end{equation}

A word of caution is in order. We can now express the action in momentum space as 
\begin{equation}
\mathcal{S}^{(NC)}_\text{int} = \frac{\lambda}{n!}\int \left(\prod_{i}^{n} \text{d}^4k_i \right)\delta\left(k_1 + \dots + k_n\right)\phi(k_1)\phi(k_2)\dots\phi(k_n)\exp\left(\frac{i}{2} \sum\limits_{i<j}^{n}k_i^\mu k_j^\nu \theta_{\mu\nu}\right),
\end{equation}
and so --- as good effective field theorists --- we may be tempted to expand the exponential for small momenta $\sim \left|k^2\right|\left|\theta\right| \ll 1$. Indeed, doing so would give us a series of irrelevant operators which would correct the leading interaction. However, once the theory is truncated at some finite order in $\theta$, we are left with a perfectly local EFT. In other scenarios where an infinite series of operators appears, this is a valid approximation procedure and allows one to calculate the leading corrections a theory predicts. But here our definition of NCFT introduces UV/IR mixing which we expect to violate EFT expectations. Truncating the series removes these effects entirely, and a theory so defined no longer has anything to do with NCFT --- at least not in the effects we will be interested in, which are nonperturbative in $\theta$ as we shall see explicitly in the following sections. There has been much work expended on these `noncommutative-inspired' theories, but they do not contain UV/IR mixing, and do not capture the most striking and most interesting features of physics on a noncommutative space, from our perspective.\footnote{We are not the first to issue a warning of this sort --- see e.g. \cite{AmelinoCamelia:2002au,Khoze:2004zc} in the context of connecting noncommutativity to the real world, and \cite{Tomboulis:2015gfa} which discusses the general case of nonlocal interactions.}

With that in mind, we may now proceed to do perturbative quantum field theory calculations, but we must worry about keeping track of all the phases from each of the vertices. In fact there is another simplification that occurs, as found by Filk \cite{Filk:1996dm}, which allows us to simplify the process of finding the phase factor for a diagram to a graph-topological statement. Filk proved two simple rules for the phase factors:
\begin{enumerate}
	\item An internal line which ends on two different vertices can be contracted while keeping the ordering of the other lines fixed. 
	\begin{equation}
	\tilde{V}\left(k_1,\dots,k_{n_1},p\right)\tilde{V}\left(-p,k_{n_1+1},\dots,k_{n_2}\right) = \tilde{V}\left(k_1,\dots,k_{n_2}\right)\delta(k_1 + \dots + k_{n_1} + p)
	\end{equation} 
	\item A loop which doesn't cross any lines can be eliminated. Note that the fixed ordering of the lines at a vertex means that we can now meaningfully speak of lines which do or don't cross each other. 
	\begin{equation}
	\tilde{V}\left(k_1,\dots,k_{n_1},p,k_{n_1+1},\dots,k_{n_2},-p\right) = \tilde{V}\left(k_1,\dots,k_{n_1},k_{n_1 + 1},\dots,k_{n_2}\right) \quad \text{if } \sum\limits_{i = n_1 + 1}^{n_2}k_i = 0
	\end{equation}
\end{enumerate}

The proof of these facts relies only on the antisymmetry of $\theta^{\mu\nu}$ and the fact that each vertex contains a momentum-conserving delta function. We may make use of this to simply find the phase factor of any Feynman diagram. Using the first rule, we can reduce any diagram to a single vertex, which is a rosette of the external lines and closed loops. The second rule allows us to eliminate loops which don't cross other lines. 

If the graph was planar (including, importantly, any tree-level graph), then by definition all loops can be eliminated. So all contributions to phase factors from internal lines cancel, and we're only left with an overall phase corresponding to the ordering of the external lines, which has remained fixed throughout the reduction process.

For a nonplanar graph, in this representation it is easy to see that we only pick up phase factors from lines which cross. The loop gives vertex legs with $\pm p^\mu$, and for an external line which doesn't cross this loop, both loop legs will be on the same side of it in the cyclic ordering, and so the two terms will cancel in the sum. Only for an external line which crosses it are the $\pm p$ on different sides, and so the antisymmetry of $\theta$ will make the two negative signs cancel to give a coherent phase for this vertex. Thus we define $I_{ij}$, the intersection matrix of an oriented graph: 
\begin{equation}
\begin{split}
I_{ij} = \begin{cases}
1& \text{line \emph{j} crosses \emph{i} from right} \\
-1 & \text{line \emph{j} crosses \emph{i} from left} \\
0 & \text{line \emph{j} does not cross \emph{i}}
\end{cases}
\end{split}
\end{equation}

\noindent Then for any graph $\mathcal{G}$, the contribution $\Gamma(\mathcal{G})$ of the phase factors is just 
\begin{equation}
\Gamma(\mathcal{G}) = \tilde{V}\left(\left\lbrace\text{external momenta}\right\rbrace\right)\times\exp\left(\frac{i}{2} \sum\limits_{ij}I_{ij} k_i \wedge k_j \right),
\end{equation}
where we've defined $k_i \wedge k_j \equiv k_i^\mu \theta_{\mu\nu}  k_j^\nu$.

In what follows we will omit the overall external phase when evaluating diagrams, as it will not be important for our purposes. We have now simplified perturbative field theory on noncommutative spaces down to the simple task of marking line-crossings, at least at the level of writing down integrands of amplitudes. The triviality of this task for tree-level graphs leads to the interesting feature that tree-level amplitudes on noncommutative spaces are the same as on commutative manifolds, and it is only at loop-level that we find deviations. We will see in the next section that the loop integration will bring surprising features. 

An important issue for the interpretation of NCFTs is that of their unitarity. There is no problem in Euclidean space, 
but for Lorentzian spacetimes with noncommutativity in the time directions (`timelike' or `space-time' noncommutativity when $- k^\mu \theta_{\mu\rho} \theta^{\rho \nu} k_\nu \equiv k \circ k < 0$ is allowed), one may find a breakdown of unitarity by taking cuts of one-loop diagrams \cite{Gomis:2000zz,Bassetto:2001vf}.\footnote{Though it is interesting to note that the special case of `lightlike' noncommutativity is also unitary \cite{Aharony:2000gz,SheikhJabbari:2010nc}.} This may be interpreted physically as being due to the production of tachyonic states, which if added to the Fock space of the theory result in a formal restoration of the cutting relations whilst making the nonunitarity explicit \cite{AlvarezGaume:2001ka}. 

This failure of unitarity is well-understood from the stringy perspective. Spatial noncommutativity appears from a background magnetic field and the field theory limit to a spacelike NCFT is smooth \cite{Seiberg:1999vs}. In the case of timelike noncommutativity, however, approaching the field theory limit forces an electric field to supercritical values whence pair-production of charged strings destabilizes the vacuum \cite{Seiberg:2000ms}. Study of string theories with timelike noncommutativity (e.g. `noncommutative open string theory' \cite{Seiberg:2000ms,Gopakumar:2000na}) is outside our scope, but there are at least some hints of similar UV/IR mixing effects as those in the NCFT \cite{Torrielli:2002ev}. We note in passing that there are further interesting connections between NCFTs and string theories --- not only do particles on noncommutative spaces act in many ways like rods of size $L \sim p \theta$ (see e.g. \cite{SheikhJabbari:1999vm,Bigatti:1999iz,Seiberg:2000gc,Girotti:2001dh,Acatrinei:2002sb}), mimicking the behavior of extended objects, but there have been many hints in the spacelike theories that the curious IR effects in the NCFT are reproducing effects from closed strings, despite the fact that these have been decoupled (e.g. \cite{Minwalla:1999px,Arcioni:2000bz,Rajaraman:2000dw,Fischler:2000fv,VanRaamsdonk:2000rr,Kiem:2000wt,Armoni:2001uw,Torrielli:2002ev,Armoni:2003va,Lopez:2003uq}).

Within the realm of field theory, there have long been suggestions that this difficulty is pointing to the need for a modified definition of quantum field theories on timelike noncommutative spaces (for some early references, see \cite{Gomis:2000gy,Bahns:2002vm,Bozkaya:2002at,Liao2002,Rim:2002if,Denk:2003jj,Fischer:2003jh,Liao:2004sw}). From this perspective, the issue is that such field theories are non-local in time, which renders nonsensical the normal time-ordering involved in the perturbative Dyson series (at the least). That is, our effective definition of these theories above via the diagrammatic expansion may be too na\"{i}ve. An interesting line of work is to formulate a modification of the standard quantum field theory machinery to non-local-in-time theories which avoids the unitarity issue by construction. We note that the same UV/IR mixing effects of interest in the two-point function have been seen to persist in at least some of these approaches (e.g. \cite{Bozkaya:2002at}). For some recent work on the formulation and properties of nonlocal field theories, see e.g. \cite{Barnaby:2007ve,Salminen:2011ut,Biswas:2014yia,Tomboulis:2015gfa, Addazi:2015dxa,Chin:2018puw}. 

Below we will begin in Euclidean space, where $k \circ k \geq 0$ is guaranteed for any $\theta_{\mu\nu}$, but will then venture into Lorentzian signature. All of our calculations and the general features we find, including finding new infrared poles, will hold robustly in spacelike noncommutative theories. However we will comment also on how these features are modified when timelike noncommutativity is turned on, taking license from the aforementioned hints that unitary completions/reformulations of timelike NCFT may retain the UV/IR mixing exhibited in the na\"{i}ve approach. 

%% file: phi4_v5.tex
In this section we review the perturbative physics of the noncommutative real scalar $\phi^4$ theory at one loop, which was first studied in detail by Minwalla, Van Raamsdonk, and Seiberg in \cite{Minwalla:1999px}.\footnote{Some early results in this model may also be found in \cite{Arefeva:1999gex,Chepelev:1999tt}.}

In four Euclidean dimensions the action on noncommutative space becomes
\begin{equation}
S = \int \text{d}^4x \left(\half \partial_\mu \phi \partial^\mu \phi + \half m^2 \phi^2 + \frac{g^2}{4!} \phi\star\phi\star\phi\star\phi\right),
\end{equation}
where we have already used the fact that the quadratic part of the noncommutative action is the same as the commutative theory to eliminate the star product there.  Our object of interest will be the one-loop correction to the two-point function. In the commutative theory this is given by a single Feynman diagram, but the noncommutative theory contains both a planar diagram and a nonplanar diagram. 

\begin{equation*}
- \Gamma^{(2)}_1 = \includegraphics[width=0.25\linewidth,valign=c]{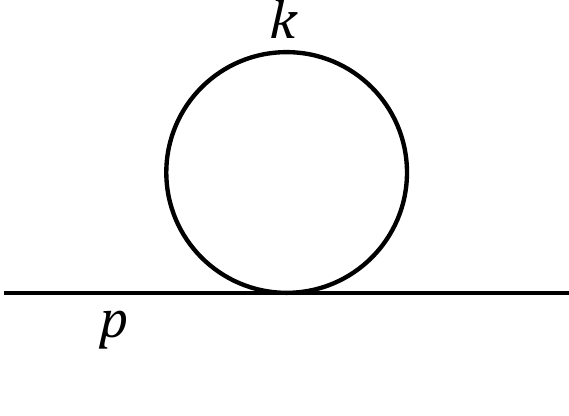} + \includegraphics[width=0.25\linewidth,valign=c]{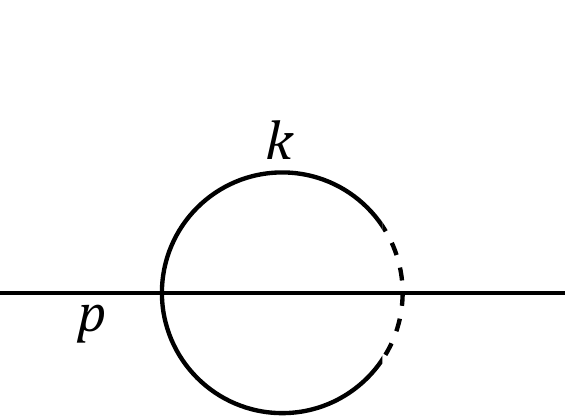}
\end{equation*} 

\noindent The expressions for these two diagrams now differ --- not only in symmetry factor but also due to the phase in the integrand. We find
\begin{align}
\begin{split}
\Gamma^{(2)}_{1,\text{planar}} &= \frac{g^2}{3\left(2\pi\right)^4} \int \frac{\text{d}^4k}{k^2 + m^2} \\
\Gamma^{(2)}_{1,\text{nonplanar}} &= \frac{g^2}{6\left(2\pi\right)^4} \int \frac{\text{d}^4k}{k^2 + m^2} e^{i k^\mu \theta_{\mu\nu} p^\nu}.
\end{split}
\end{align}
We may already see that something interesting should happen, as in the nonplanar diagram the phase mixes the internal and external momenta. One may intuit that the rapidly oscillating phase in the UV of the loop integration will dampen the would-be divergence, and indeed we will see that nonplanar diagrams are finite. However, unlike in the case where the vertex factor vanishes rapidly for large Euclidean momenta and so ensures UV-finiteness \cite{Chin:2018puw}, here the damping is in some sense `marginal'. This fact will be responsible for the interesting feature we will find presently. 

The simplest method to evaluate noncommutative diagrams is to use Schwinger parameters, recalling the identity $\frac{1}{k^2+m^2} = \int_{0}^{\infty}\text{d}\alpha \ e^{-\alpha\left(k^2 + m^2\right)}$. The presence of the phase in the nonplanar diagram means we must complete the square before going to spherical coordinates to get a Gaussian integral. This means that after the momentum integrals we end up with 
\begin{align}\label{eqn:phi4gaussianint}
\begin{split}
\Gamma^{(2)}_{1,\text{planar}} &= \frac{g^2}{48\pi^2} \int \frac{\text{d}\alpha}{\alpha^2} e^{-\alpha m^2} \\
\Gamma^{(2)}_{1,\text{nonplanar}} &= \frac{g^2}{96\pi^2} \int \frac{\text{d}\alpha}{\alpha^2} e^{-\alpha m^2-\frac{p\circ p}{4\alpha}}
\end{split}
\end{align}
where again $p \circ q = - p^\mu\theta^2_{\mu\nu} q^\nu$. Moving to Schwinger space trades large-$k$ divergences for small-$\alpha$ divergences, which we now smoothly regulate by multiplying the integrands by $\exp\left(-1/(\Lambda^2 \alpha)\right)$ so that the small $\alpha$ region will be driven to zero. Note that a term of this form already exists in the expression for the nonplanar diagram. After introducing the regulator, we can evaluate the integrals to find 
\begin{align}
\begin{split}
\Gamma^{(2)}_{1,\text{planar}} &= \frac{g^2}{48\pi^2} \left(\Lambda^2 - m^2 \log\left(\frac{\Lambda^2}{m^2}\right) + \mathcal{O}(1)\right) \\
\Gamma^{(2)}_{1,\text{nonplanar}} &= \frac{g^2}{96\pi^2} \left(\Lambda_\text{eff}^2 - m^2 \log\left(\frac{\Lambda_\text{eff}^2}{m^2}\right) + \mathcal{O}(1)\right),
\end{split}
\end{align}
where we've defined 
\begin{equation}
\Lambda_\text{eff}^2 \equiv \frac{1}{1/\Lambda^2 + p \circ p/4},
\end{equation}
which is the effective cutoff of the nonplanar diagram. 

The first thing to note is that it seems the UV divergence of the nonplanar diagram has disappeared --- the graph is finite in the limit $\Lambda\rightarrow \infty$, and so appears to have been regulated by the noncommutativity of spacetime. In fact the effect is more subtle, as alluded to earlier, and now the UV and IR limits of this amplitude do not commute. If we first take an infrared limit $p\circ p \rightarrow 0$ we find that $\Lambda_{\text{eff}} \rightarrow \Lambda$ and the ultraviolet divergence of the commutative theory reappears. If we take the UV limit $\Lambda \rightarrow \infty$ first we find an IR divergence $\frac{1}{p \circ p}$, so the noncommutativity has transmogrified the UV divergence into an IR one.\footnote{We note here that the failure of a `correspondence principle' between commutative and noncommutative theories as $\theta^{\mu\nu} \rightarrow 0$ is clearly intrinsically linked to the appearance of UV/IR mixing. This failure doesn't violate Kontsevich's proof of the existence of deformation quantization for any symplectic manifold \cite{Kontsevich:1997vb}, as that is confined solely to `formal' deformation quantization --- that is, the production of a formal power series expansion of the algebra of observables in terms of the deformation parameter. As was noted in Section \ref{sec:review} and is now on prime display, the physics of the theory with nonperturbative $\theta$-dependence is starkly different from that of any truncation.}

Turning to the question of renormalizability, one may na\"{i}vely ask if we can absorb all UV divergences into a finite number of counterterms. Under this criterion, it is clear that this procedure works in the noncommutative theory at least when the commutative version is renormalizable. In the current case, we may absorb the UV divergences of this correction to the two-point function into a redefinition of the physical mass, $M^2 = m^2 + \frac{g^2 \Lambda^2}{48 \pi^2} - \frac{g^2 m^2}{48 \pi^2}\log\frac{\Lambda^2}{m^2}$, and so write down a one-particle irreducible quadratic effective action which has a finite $\Lambda\rightarrow\infty$ limit: 
\begin{equation} \label{eqn:1PIaction}
S^{(2)}_{1\text{PI}} = \int \frac{\text{d}^4p}{(2 \pi)^4} \ \half \left(p^2 + M^2 + \frac{g^2}{96\pi^2\left(\frac{p\circ p}{4} + \frac{1}{\Lambda^2}\right)} - \frac{g^2 M^2}{96 \pi^2} \log \frac{1}{M^2\left(\frac{p \circ p}{4} + \frac{1}{\Lambda^2}\right)} + \dots + \mathcal{O}(g^4)\right)\phi(p)\phi(-p).
\end{equation}

However, in the $\Lambda \rightarrow \infty$ limit one finds that at one loop the propagator now has \textit{two} poles. The first is a standard radiative correction to the free pole, but the second has appeared ex nihilo at one loop:
\begin{align}
\begin{split}
p^2 &= - m^2 + \mathcal{O}(g^2) \\
p \circ p &= - \frac{g^2}{24 \pi^2 m^2} + \mathcal{O}(g^4),
\end{split}
\end{align}
where we have assumed that $\theta^{\mu\nu}$ is full rank. The former is to be interpreted as the on-shell propagation of the particles associated to our fundamental field $\phi$. If $\theta^{\mu\nu}$ has only one eigenvalue $1/\Lambda_\theta^2$ --- with $\Lambda_\theta$ thought of as the scale associated with the breakdown of classical geometry --- we have $p \circ p = \frac{p^2}{\Lambda_\theta^4}$. We see that the new pole appears at $p^2 \propto g^2 \frac{\Lambda_\theta^4}{m^2}$, and so if our field $\phi$ lives in the deep UV of the theory, our new pole appears at parametrically low energy scales. To the extent that poles are particles, we appear to have generated a new light particle from ultraviolet dynamics.\footnote{Although there is no pole at finite $\Lambda$, a scale is still induced in the form of an infrared cutoff $\sim \Lambda_\theta^2/\Lambda$.}

The interpretation of the new pole can be sharpened by considering more carefully the criteria for renormalizability in Wilsonian EFT. In a Wilsonian picture, we upgrade our Lagrangian parameters to running parameters, and define our theory at the scale $\Lambda$ as 
\begin{equation}
S_{Wilson}(\Lambda) = \int \text{d}^4x \left(\half Z(\Lambda) \partial_\mu \phi \partial^\mu \phi + \half Z(\Lambda) m^2(\Lambda) \phi^2 + \frac{Z^2(\Lambda) g^2(\Lambda)}{4!} \phi\star\phi\star\phi\star\phi\right).
\end{equation}
It is immediately apparent from the above calculation that we cannot write the action at a lower scale $\Lambda_0 < \Lambda$ in this same form by choosing appropriate definitions for $Z(\Lambda),m(\Lambda),g(\Lambda)$ --- there's nowhere to put the $\frac{1}{p \circ p}$ term!\footnote{There has been much work on understanding renormalizability of NCFTs, especially with an eye toward finding a mathematically well-defined four-dimensional quantum field theory with a non-trivial continuum limit. Renormalizability has been proven for modifications of NCFTs where the free action is supplemented by an additional term which adjusts its long-distance behavior. Such an action is manufactured either by requiring it manifest `Langman-Szabo' duality \cite{Langmann:2002cc} $p_\mu \leftrightarrow 2 (\theta^{-1})_{\mu\nu}x^\nu$ \cite{Grosse:2004yu,Grosse:2012uv} or by adding a $1/p\circ p$ term to the free Lagrangian \cite{Gurau:2008vd}, the latter of which directly has the interpretation of adding `somewhere to put the $1/p\circ p$ counterterm'. For recent reviews of these and related efforts we refer the reader to \cite{Grosse:2016yjo,Ydri:2016dmy}. It would be interesting to understand fully the extent to which the physics of these schemes agrees with the interpretation of the IR effects as coming from auxiliary fields \cite{Minwalla:1999px,VanRaamsdonk:2000rr}.}	

Stated more precisely, for Wilsonian renormalizability we require that we can define the running couplings such that correlation functions computed from this action converge uniformly to their $\Lambda \rightarrow \infty$ limits. However, this requirement is flatly violated by the noncommutation of the UV and IR limits of the diagrams. For any finite value of $\Lambda$, the effective action of Equation \ref{eqn:1PIaction} differs significantly from its limiting value for small momenta $p \circ p \ll \frac{1}{\Lambda^2}$. This is the precise sense in which the violation of Wilsonian EFT appears in this one-loop correction.

This brings up the question of how an effective field theorist would describe the universe if they unknowingly lived on a noncommutative space. A consistent Wilsonian interpretation can be regained by including a degree of freedom which can absorb the new infrared dynamics of the quadratic effective action. Since we need this to involve the $\phi$ momentum, this new particle must mix linearly with the $\phi$ field. We manufacture its tree-level Lagrangian such that the problematic inverse $p \circ p$ term in the quadratic effective action of $\phi$ is replaced with its $\Lambda \rightarrow \infty$ value for all values of $\Lambda$, to satisfy our precise condition for Wilsonian renormalizability. To see how this works, we add to our tree level Wilsonian action
\begin{equation}\label{eqn:auxaction}
\Delta S(\Lambda) = \int \text{d}^4x \left(\half \partial\chi \circ \partial\chi + \half \frac{\Lambda^2}{4} (\partial \circ \partial \chi)^2 + i \frac{1}{\sqrt{24 \pi^2}} g \chi \phi\right).
\end{equation} 
Since $\chi$ appears quadratically, we may integrate it out exactly at tree level to find a contribution to the effective action 
\begin{equation}
\Delta S_{1\text{PI}}(\Lambda) = \int \frac{\text{d}^4p}{(2\pi)^4} \half \left(-\frac{g^2}{96\pi^2\left(\frac{p\circ p}{4} + \frac{1}{\Lambda^2}\right)} + \frac{g^2}{24\pi^2 p\circ p}\right)\phi(p)\phi(-p)
\end{equation}

This precisely subtracts off the problematic term in the original 1PI quadratic effective action and adds back its $\Lambda \rightarrow \infty$ limit, as we had wanted. Ignoring the logarithmic term,\footnote{Discussion of the interpretation of logarithmic singularities as being due to auxiliary fields propagating in extra dimensions may be found in \cite{VanRaamsdonk:2000rr}.} we are left with an effective action which is manifestly independent of the cutoff $\Lambda$, and so satisfies our criterion for Wilsonian renormalizability.\footnote{In Equation \ref{eqn:auxaction}, the four-derivative quadratic action of the auxiliary field can be rewritten as two fields with two-derivative actions, one of which is of negative norm and may be thought of as the `Lee-Wick partner' of the positive norm state \cite{Lee:1969fy}, viz.
\begin{equation}
\mathcal{L} = \half \partial \chi' \circ \partial \chi' - \half \partial \tilde{\chi} \circ \partial \tilde{\chi} - \half \frac{4}{\Lambda^2} \tilde{\chi}^2 + i \frac{1}{\sqrt{24 \pi^2}} g\left(\chi' - \tilde{\chi}\right)\phi , \qquad \chi' \equiv \chi + \tilde{\chi} 
\end{equation}
One may then wonder if the lightness of the new IR pole may be understood through the regularization performed by the Lee-Wick field, as is done for the Higgs in the `Lee-Wick standard model' \cite{Grinstein:2007mp}. However, in that theory the Higgs is kept light because every particle comes with a Lee-Wick partner, and so all diagrams contributing to corrections to the Higgs mass are made finite. The presence of the Higgs' Lee-Wick partner alone is not enough to keep it light. Here, the lightness of $\chi$ can be understood diagrammatically as being simply due to the fact that its only interaction is linear mixing with $\phi$, and so any correction to its two-point function is absorbed into that of the two-point function of $\phi$. A further issue with the Lee-Wick rewriting is that the seeming perturbative unitary of the theory is normally guaranteed by the Lee-Wick partner being heavy and unstable. But as we take the $\Lambda\rightarrow\infty$ limit in our Wilsonian action, we see that the Lee-Wick partner becomes massless as well, in accordance with the result that this theory is non-unitary \cite{Gomis:2000zz}.} We discuss the generalization of this procedure in Appendix \ref{app:auxfield}.

Now while we have written down an action which identifies the new observed IR pole with a field and in doing so gives our effective action a Wilsonian interpretation, the extent to which $\chi$ can be taken seriously as a fundamental degree of freedom is unclear.\footnote{We note that in matrix models containing dynamical noncommutative geometries it has been argued that emergent infrared singularities should be associated with the dynamics of the geometry (see e.g. \cite{VanRaamsdonk:2001jd,Steinacker:2010rh}). As our field theories are formulated on fixed noncommutative backgrounds, this interpretation is unavailable to us.} The new pole is inaccessible in Euclidean space --- so one does not immediately conclude there is a tachyonic instability --- and relatedly, when we na\"{i}vely analytically continue this result to Lorentzian spacetime this new pole is inaccessible in the $s$-channel.\footnote{Note that this peculiar connection regarding (in)accessibility is due to the Lorentz violation.   While the normal pole which is inaccessible in Euclidean signature becomes accessible for timelike momenta in Lorentzian signature, the Wick rotation affects the noncommutative momentum contraction differently. When taking $x_4 \rightarrow - i x_0$, one also rotates $\theta_{4\nu} \rightarrow -i \theta_{0\nu}$ such that Equation \ref{eqn:ncdef} continues to hold for the same numerical $\theta_{\mu\nu}$. For the simplest configuration of full-rank noncommutativity with $\theta_{\mu\nu}$ block-off-diagonal and only one eigenvalue $1/\Lambda_\theta^2$, the Euclidean $p \circ p = p^2/\Lambda_\theta^4$ becomes a Lorentzian $p \circ p = (p_0^2 - p_1^2 + p_2^2 + p_3^2)/\Lambda_\theta^4$. So a noncommutative pole which is inaccessible in the Euclidean theory becomes accessible in the Lorentzian theory for spacelike momenta, while a noncommutative pole which can be accessed in the Euclidean theory becomes accessible in the $s$-channel in Lorentzian signature.} However, its presence is still enough to break unitarity for this theory \cite{Gomis:2000zz}, and in fact may still be interpreted as being due to the presence of tachyons \cite{AlvarezGaume:2001ka}. As discussed in Section \ref{sec:review}, it is possible this may be resolved if analytical continuation is adjusted for nonlocal-in-time theories, or it may be that a UV theory cures this apparent violation. 

Separately, it is not obvious much has been gained by attributing the new pole to a new, independent field, past acting as a formal tool to regain a notion of renormalizability. Since the only interaction of $\chi$ above is linear mixing, its action is not renormalized --- any divergences are instead absorbed into the running of $\phi$ parameters --- and so no interactions are generated. Furthermore one is obstructed from integrating out the heavy field $\phi$ to come up with an effective action of $\chi$ at low energies by the fact that the kinetic terms of $\chi$ are non-standard, which prevents diagonalization of the quadratic terms in the Lagrangian. Thus it seems it is intrinsically linked with the heavy scalar which begat it.


There are further obstructions to asking that this specific mechanism be responsible for the lightness of an observed particle such as the Higgs. Prime among these is the modified dispersion relation of the new field, $p \circ p = \mathcal{O}(g^2)$, which means that the free propagation of this field would be Lorentz violating.\footnote{This dispersion relation means that $\chi$ only propagates in noncommutative directions, and so attempts to use hidden extra-dimensional noncommutativity to avoid four-dimensional Lorentz violation constraints seem a phenomenological nonstarter.} We will explore these issues further in the next sections, as in the Yukawa theory of Section \ref{sec:yukawa} the new pole will appear with the opposite sign and so will offer the prospect of appearing as an $s$-channel pole. 

We emphasize that a new infrared scale whose lightness is unexplained in the context of Wilsonian effective field theory is an exciting feature that makes further exploration of UV/IR mixing an interesting pursuit. The fact that it here appears as the scale of a pole in a propagator makes the connection to the hierarchy problem captivating, but asking that this toy model --- where Lorentz violation is at the fore --- literally solve the problem for us would be too much. We proceed without further hindrance in exploring NCFT so as to learn more about the appearance and effects of UV/IR mixing here.

\subsection{Dimensional Regularization} \label{sec:dimreg}

A good question to ask is whether, or to what extent, these effects are an artifact of our choice of regularization. To demonstrate their physicality, we repeat the calculation of the one-loop correction to the two-point function now in dimensional regularization. We set up our integral in $d=4-\epsilon$ dimensions, having defined $g^2 = \tilde{g}^2 \tilde{\mu}^\epsilon$, and we again go to Schwinger space:
\begin{align}
\begin{split}
\Gamma^{(2)}_{1,\text{planar}} &= \frac{\tilde{g}^2 \tilde{\mu}^\epsilon}{3\left(2\pi\right)^d} \int \text{d}^dk \ \text{d}\alpha \ e^{-\alpha(k^2 + m^2)} \\
\Gamma^{(2)}_{1,\text{nonplanar}} &= \frac{\tilde{g}^2 \tilde{\mu}^\epsilon}{6\left(2\pi\right)^d} \int \text{d}^dk \ \text{d}\alpha \ e^{-\alpha(k^2 + m^2) + i k^\mu \theta_{\mu\nu} p^\nu}.
\end{split}
\end{align}

After completing the square in the nonplanar integral, the momentum integral and the Schwinger integral may then be performed analytically, with the results: 
\begin{align} \label{eqn:dimregint}
\begin{split}
\Gamma^{(2)}_{1,\text{planar}} &= \frac{\tilde{g}^2 \tilde{\mu}^\epsilon}{3\left(4\pi\right)^{d/2}} (m^2)^{\frac{d}{2}-1} \Gamma(1 - \frac{d}{2}) \\
\Gamma^{(2)}_{1,\text{nonplanar}} &= \frac{\tilde{g}^2 \tilde{\mu}^\epsilon}{6 \left(4\pi\right)^{d/2}} 2^{\frac{d}{2}} (m^2)^{\half({\frac{d}{2}-1})} \left(\sqrt{p \circ p}\right)^{1 - \frac{d}{2}} K_{\frac{d}{2}-1}\left(m\sqrt{p\circ p}\right).
\end{split}
\end{align}
If we expand the planar graph in the limit $\epsilon \rightarrow 0$, which should be thought of as probing the ultraviolet, we recover
\begin{equation}
\Gamma^{(2)}_{1,\text{planar}} = - \frac{\tilde{g}^2 m^2}{3(4 \pi)^2} \left[ \frac{2}{\epsilon} + \ln\frac{\mu^2}{m^2}\right],
\end{equation}
where in $\overline{\text{MS}}$ we would subtract off the pole and find the renormalization group evolution of $m$ from the logarithmic term, as usual. 

The question of dimensional regularization for the nonplanar diagram is a subtle one \cite{Huffel:2002pv}. If we first take the $\epsilon \rightarrow 0$ limit of Equation \ref{eqn:dimregint}, we see this manifestly has no divergences, and we are simply left with the finite, $\epsilon^0$ term
\begin{equation} \label{eqn:dimregepsfirst}
\Gamma^{(2)}_{1,\text{nonplanar}} = \frac{g^2 m^2}{6(4\pi)^2}\left[\frac{4}{m^2 p\circ p}- \ln \frac{4}{m^2 p \circ p} - 1 + 2 \gamma \right],
\end{equation}
which we have expanded near $p \circ p \rightarrow 0$ to manifest the IR divergence. We have again transmogrified our UV divergence into an IR pole. We now expect to see that the IR limit does not commute with the above UV limit. To do so, we expand Equation \ref{eqn:dimregint} around $p \circ p \rightarrow 0$ to find
\begin{equation}\label{eqn:dimregpopfirst}
\Gamma^{(2)}_{1,\text{nonplanar}} = \frac{\tilde{g}^2 m^2}{6 (4\pi)^2} \frac{\pi^{\epsilon/2}\tilde{\mu}^\epsilon}{m^\epsilon} \Gamma\left(-1 + \frac{\epsilon}{2}\right) + \frac{\tilde{g}^2}{24 \pi^2}\tilde{\mu}^\epsilon \pi^{\epsilon/2} \Gamma\left(1 - \frac{\epsilon}{2}\right) p \circ p^{-1 + \epsilon/2} + \mathcal{O}(p \circ p).
\end{equation}

If we were to now blindly take the $\epsilon\rightarrow 0$ limit of this expression, we would again get Equation \ref{eqn:dimregepsfirst}, contrary to our expectations. However, we notice that if the dimension of spacetime over which we had performed the integral was particularly low $\epsilon > 2$, then we have incorrectly kept the second term in Equation \ref{eqn:dimregpopfirst}, as that term would be at least $\mathcal{O}(p \circ p)$. If we were to work in $d < 2$, expand in $p \circ p \rightarrow 0$ and so ignore that term, and \textit{then} analytically continue back to $d = 4$, we would instead find the $\epsilon^{-1}$ pole
\begin{equation} \label{eqn:dimregnppole}
\Gamma^{(2)}_{1,\text{nonplanar}} = - \frac{\tilde{g}^2 m^2}{6(4 \pi)^2} \left[ \frac{2}{\epsilon} + \ln\frac{\mu^2}{m^2}\right],
\end{equation}
and now we recover the UV divergence that was present in the commutative theory, so that once again we find the UV and IR limits don't commute.

The key to understanding clearly this seemingly ambiguous dimensional regularization procedure is that while  $\Gamma^{(2)}_{1,\text{nonplanar}}(p \circ p) \sim \int \text{d}^dq \ \text{d}\alpha \ e^{-\alpha(q^2 + m^2) - \frac{p\circ p}{4 \alpha}}$ is convergent in $d>2$ for $p \circ p > 0$, at $p \circ p = 0$ it is only convergent for $d < 2$. Since it is a property of dimensional regularization that if an integral converges in $\delta$ dimensions, it converges to the same value in $d < \delta$ dimensions \cite{Collins:1984xc}, we may thus perform the integral at $d < 2$ for \textit{all} $p \circ p$ and correctly find Equation \ref{eqn:dimregint}. It is only when taking the IR limit that we must remember the integral was performed in $d < 2$ dimensions, and so our expansion to get Equation \ref{eqn:dimregnppole} is unambiguously correct. Thus our conclusion that the UV and IR limits of the two-point function do not commute here is robust. 

It is thus clear that the UV/IR mixing we have observed in this model is not an artifact of a choice of regularization, and is in fact a physical feature of this noncommutative field theory.

%% file: yukawa_v5.tex
\subsection{Motivation: Strong UV/IR Duality}

We observed in our first example that the UV divergences of the real $\phi^4$ commutative theory are transmogrified into infrared poles in the noncommutative theory.\footnote{While we only presented the calculation of the one-loop correction to the two-point function, \cite{Minwalla:1999px} goes through corrections to the two- and $n$-point functions for $\phi^n$ with $n=3,4$ and finds the same features in all cases.} It is natural to ask whether this ``strong UV/IR duality'' \cite{RuizRuiz:2002hh} is a common feature of \textit{all} noncommutative theories.

The answer is no, and the simplest counterexample is provided in the case of a complex scalar field with global $U(1)$ symmetry and self-interaction \cite{RuizRuiz:2002hh}. In the quantization of the scalar potential we have two quartic terms which are noncommutatively-inequivalent due to the ordering non-invariance, so the general noncommutative potential is 
\begin{equation}\label{eqn:complexphi4}
V = m^2 |\phi|^2 + \frac{\lambda_1}{4} \phi^* \star \phi \star \phi^* \star \phi + \frac{\lambda_2}{4} \phi^* \star \phi^* \star \phi \star \phi,
\end{equation}
where $\lambda_1$ and $\lambda_2$ are now different couplings. By doodling some directed graphs, one sees simply that the one-loop correction to the scalar two-point function contains planar graphs with each of the $\lambda_1, \lambda_2$ vertices, but the only nonplanar graph has a $\lambda_2$ vertex. There is thus no necessary connection of the ensuing nonplanar IR singularity to the UV divergence in the $\theta\rightarrow 0$ limit, as the coefficients are unrelated (and in particular, we are free to turn off the IR singularity at one loop by setting $\lambda_2 = 0$).

Another important counterexample is that of charged scalars, the simplest example of which is noncommutative scalar QED, which was first constructed in \cite{Arefeva:2000vow}. There is a very rich and interesting structure of gauge theories on noncommutative spaces, a full discussion of which is far beyond the scope of this paper. We refer the reader to \cite{Armoni:2000xr,Chaichian:2001py,Chaichian:2001mu,Chaichian:2004yw,Khoze:2004zc,Arai:2007dm} for discussions of some features relevant to SM model-building. We here satisfy ourselves with the simplest case, for which we have the noncommutative Lagrangian\footnote{It is important to note that many fundamental concepts which one normally thinks of as depending upon Lorentz invariance still hold on noncommutative spaces, due to a `twisted Poincar\'{e} symmetry' \cite{Oeckl:2000eg,Wess:2003da,Chaichian:2004yh,Chaichian:2004za}. This includes the unitary irreducible representations, so it is sensible to speak of a vector field.} 
\begin{equation}
\mathcal{L} = \frac{1}{4g^2} F_{\mu\nu} \star F^{\mu\nu} + (D_\mu \phi)^* \star (D^\mu \phi) + V\left(\phi,\phi^*\right),
\end{equation}
where even though we're quantizing $U(1)$ we have $F_{\mu\nu} = \partial_\mu A_\nu - \partial_\nu A_\mu - i \left[A_\mu \stackrel{*}{,} A_\nu\right]$ due to the noncommutativity, where $\left[\cdot \stackrel{*}{,} \cdot\right]$ is the commutator in our noncommutative algebra. The vector fields transform as $A_\mu \mapsto U \star A_\mu \star U^\dagger + i \partial_\mu U \star U^\dagger$, where $U(x)$ is an element of the noncommutative $U(1)$ group, which consists of functions $U(x) = \left(e^{i\theta(x)}\right)_\star$, which is the exponential constructed via power series with the star-product.

The potential and the covariant derivative both depend on the representation we choose for the scalar. In contrast to commutative $U(1)$ gauge theory, where we merely assign $\phi$ a charge, our only choices now are to put $\phi$ in either the fundamental or the adjoint of the gauge group. Note that an adjoint field smoothly becomes uncharged in the commutative limit. Such a field $\phi$ transforms as  $\phi \mapsto U \star \phi \star U^\dagger$. The covariant derivative is thus $D^\mu \phi = \partial_\mu \phi - i g \left[A_\mu \stackrel{*}{,}\phi\right]$. The gauge-invariant potential then includes both quartic terms in Equation \ref{eqn:complexphi4}, in addition to others such as $\phi^* \star \phi \star \phi \star \phi$, since the adjoint complex scalar is uncharged at the level of the global part of the gauge symmetry. Strong UV/IR duality then should not hold here either.

The situation is even worse if $\phi$ is in the fundamental, where it transforms as $\phi \mapsto U \star \phi$ and $\phi^* \mapsto \phi^* \star U^{-1}$ with covariant derivative $D^\mu \phi = \partial^\mu \phi - i A_\mu \star \phi$. It is easy to see in this case that the $\lambda_2$ interaction term is no longer gauge invariant, and a charged scalar may only self-interact through $V = \lambda_1 \phi^* \star \phi \star \phi^* \star \phi$. Purely from gauge invariance we thus see that a fundamental scalar has no nonplanar self-interaction diagrams in the one-loop correction to its two-point function, and so there is no remnant of strong UV/IR duality to speak of.\footnote{Noncommutative QED also has strange behavior in the gauge sector that runs counter to strong UV/IR duality --- the photon self-energy correction gains an infrared singularity from nonplanar one-loop diagrams, even though the commutative quadratic power-counting divergence is forbidden by gauge-invariance. The theory is constructed in detail in \cite{Hayakawa:1999zf}, while more physical interpretation is given in \cite{Matusis:2000jf}, and the possible relation to geometric dynamics in the context of matrix models is discussed in \cite{VanRaamsdonk:2001jd}.}

The question is then whether there are other examples where this strong UV/IR duality \textit{does} occur, or whether it is perhaps a peculiar feature of real $\phi^n$ theories on noncommutative spaces. To answer this, we will study in detail another case of especial phenomenological significance: Yukawa theory. Noncommutative Yukawa theory was first studied in \cite{Anisimov:2001zc}.\footnote{Aspects of noncommutative Yukawa theory have also been studied recently in d=3 in \cite{Bufalo:2016mui}, and with a modified form of noncommuativity in \cite{Bouchachia:2015kxa}.} Our result on the presence of strong UV/IR mixing differs, for reasons we will explain henceforth.

\subsection{Setup}

For reasons that will soon become clear, we will now work directly in Minkowski space, and begin with a commutative theory of a real scalar $\varphi$ and a Dirac fermion $\psi$ with Yukawa interaction: 
\begin{equation}
\mathcal{L^{(\text{C})}} = -\half \partial_\mu \varphi \partial^\mu \varphi - \half m^2 \varphi^2 + i \overline{\psi} \slashed{\partial} \psi  - \overline{\psi} M \psi + g \varphi \overline{\psi} \psi.
\end{equation}

When constructing a noncommutative version of this theory, the quadratic part of the action does not change. However, ordering ambiguities appear for the interaction term, and we in fact find two noncommutatively-inequivalent interaction terms which generically appear:
\begin{equation}\label{eqn:yukint}
\mathcal{L^{(\text{NC})}_{\text{int}}} = g_1 \varphi \star \overline{\psi} \star \psi + g_2  \overline{\psi} \star \varphi \star \psi.
\end{equation}

\begin{figure}[!ht]
	\begin{subfigure}[b]{0.4\textwidth}
		\includegraphics[width=\textwidth]{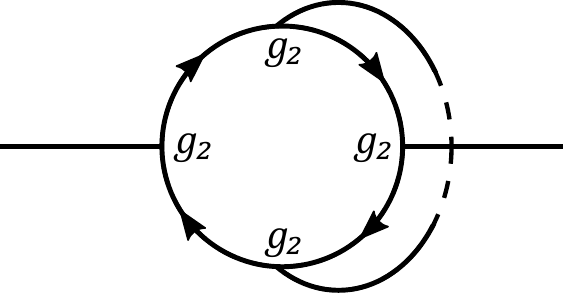}
	\end{subfigure}
	\hfill
	\begin{subfigure}[b]{0.4\textwidth}
		\includegraphics[width=\textwidth]{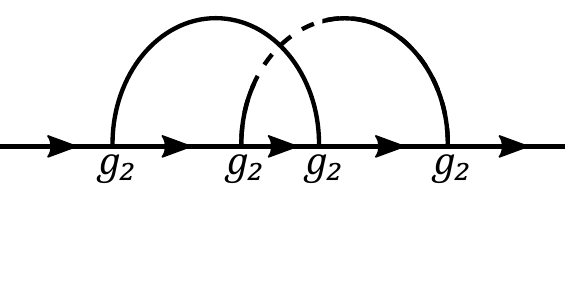}
	\end{subfigure}
	\caption{Representative leading nonplanar corrections to the self-energies in the noncommutative Yukawa theory of \cite{Anisimov:2001zc}. Fermion lines have arrows and dashing denotes nonintersection.} \label{fig:yuk2loop}
\end{figure}

These terms are inequivalent because the star product is only cyclically invariant. In the analysis of \cite{Anisimov:2001zc}, only the $g_2$ interaction was included. As a result, it was concluded that this theory contains no nonplanar diagrams at one loop, and the first appear at two loops as in Figure \ref{fig:yuk2loop}. This immediately tells us that the one-loop quadratic divergence of the scalar self-energy will not appear with a one-loop IR singularity, and so rules out the putative strong UV/IR duality of the theory they studied.

However, we must ask whether we actually have the freedom to choose $g_1$ and $g_2$ independently. To address that question, we must understand the role of discrete symmetries in noncommutative theories. For ease of reference we here repeat our definition of the noncommutativity parameter 
\begin{equation}\tag{\ref{eqn:ncdef}}
\left[x_\mu,x_\nu\right] = i \theta_{\mu\nu}
\end{equation}
It is manifest that the noncommutativity tensor does not transform homogeneously under either parity or time-reversal, but only under their product: $PT: x_\mu \rightarrow - x_\mu \Rightarrow PT: \theta_{\mu\nu} \rightarrow \theta_{\mu\nu}$. So while any Lagrangian with full-rank noncommutativity unavoidably violates both $P$ and $T$, it may preserve $PT$. 

Since both $\varphi$ and the scalar fermion bilinear are invariant under all discrete symmetries, these symmetries na\"{i}vely play no further role in this theory. However, the time-reversal operator is anti-unitary, and thus negates the phase in the star-product: 
\begin{equation}
(PT)^{-1} \left(f(x) \star g(x)\right) PT =  g(x) \star f(x).
\end{equation}

\noindent Armed with this, we may now apply CPT to our interaction Lagrangian, to find

\begin{equation}
(CPT)^{-1} \mathcal{L^{(\text{NC})}_{\text{int}}} CPT = g_1   \overline{\psi} \star\varphi \star \psi  + g_2 \varphi \star \overline{\psi} \star \psi.
\end{equation}
Comparing with Equation \ref{eqn:yukint}, we see that our interactions have been re-cycled! Requiring that our interactions preserve CPT amounts to imposing
\begin{equation}
(CPT)^{-1} \mathcal{L^{(\text{NC})}_{\text{int}}} CPT= \mathcal{L^{(\text{NC})}_{\text{int}}} \quad \Longrightarrow \quad  g_1 = g_2
\end{equation}

And so the theory of \cite{Anisimov:2001zc} appears to violate CPT.\footnote{We note that while the CPT theorem has only been proven in NCFT without space-time noncommutativity \cite{Chaichian:2002vw,Franco:2004gx,AlvarezGaume:2003mb,Soloviev:2006ah}, the difficulty in the general case is related to the issues with unitarity discussed in Section \ref{sec:review}, and we expect it should hold in a sensible formulation of the space-time case as well.} When we instead include both orderings of interactions the nonplanar diagrams now occur at the first loop order. Furthermore, with both couplings set equal the planar and nonplanar diagrams will have the same coefficients, which reopens the question of strong UV/IR duality for this theory. In the following we will keep $g_1$ and $g_2$ distinguished merely to evince how the different vertices appear, but in drawing conclusions about the theory we will set them equal.\footnote{We should note that in the construction of noncommutative QED it has been argued that it is sensible to assign $\theta$ the anomalous charge conjugation transformation $C: \theta^{\mu\nu} \rightarrow - \theta^{\mu\nu}$ (\cite{SheikhJabbari:2000vi} and many others since). The argument is that charged particles in noncommutative space act in some senses like dipoles whose dipole moment is proportional to $\theta$, and so charge conjugation should naturally reverse these dipole moments. Here, however, our particles are uncharged, and thus we have no basis for arguing in this manner. Furthermore, such an anomalous transformation makes charge conjugation relate theories on \textit{different} noncommutative spaces $\mathcal{M}_\theta \rightarrow \mathcal{M}_{-\theta}$. The heuristic picture of the CPT theorem (that is, the reason we care about CPT being a symmetry of our physical theories) is that after Wick rotating to Euclidean space, such a transformation belongs to the connected component of the Euclidean rotation group \cite{Witten:2018lha}, and so is effectively a symmetry of spacetime. So it is at the least not clear that defining a CPT transformation that takes one to a different space accords with the reason CPT should be satisfied in the first place.}

\subsection{Scalar Two-Point Function}

First we consider the planar diagrams, of which there are two: \\
\begin{equation*}
-i\Gamma^{2,s,p}_1(p) \quad = \quad  \includegraphics[width=0.25\linewidth,valign=c]{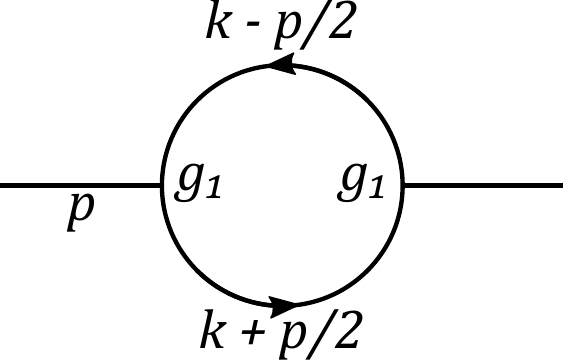} \quad + \quad \includegraphics[width=0.25\linewidth,valign=c]{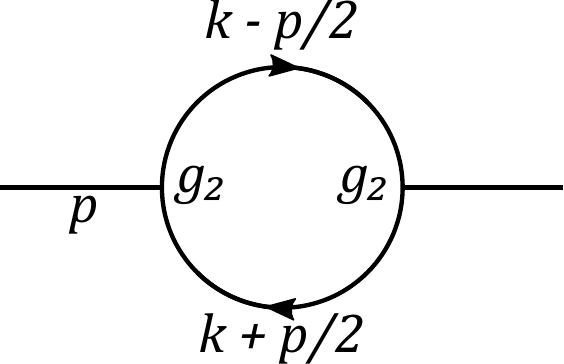}
\end{equation*} 

The `symmetrization' of the momenta of the internal propagators is an important calculational simplification. This calculation is textbook save for our Schwinger-space regularization, so we will be brief and merely point out the salient features. The sum of these diagrams gives
\begin{equation}
\Gamma^{(2),s,p}_1(p) = i(-1)\left((i g_1)^2 + (i g_2)^2\right) \int \frac{\text{d}^4k}{(2\pi)^4} \frac{(-i)^2 \text{Tr}\left[\left(M -  \slashed{k}-\slashed{p}/2\right)\left(M - \slashed{k} +  \slashed{p}/2\right)\right]}{\left((k+p/2)^2 + M^2\right)\left((k-p/2)^2 + M^2\right)}.
\end{equation}

To evaluate this, we must now introduce two Schwinger parameters $\alpha_1, \alpha_2$ and then switch to `lightcone Schwinger coordinates' which effects the change $\int_0^\infty \text{d} \alpha_1 \int_0^\infty \text{d} \alpha_2 \rightarrow \int_0^\infty \text{d} \alpha_+ \int_{-\alpha_+}^{+\alpha_+} \text{d} \alpha_-$. Regulating the integral by $\exp\left[-1/\sqrt{2} \alpha_+ \Lambda^2 \right]$, we may then evaluate and isolate the divergences as $\Lambda \rightarrow \infty$ to find

\begin{equation} \label{eqn:scalaroneloopplanar}
\Gamma^{(2),s,p}_1(p) = - \frac{(g_1^2 + g_2^2)}{2 \pi^2} \left[\Lambda^2 - \frac{6  M^2 + p^2}{4} \log\left(\frac{\Lambda^2}{M^2 + p^2/4}\right) + \dots \right]
\end{equation} 

\noindent Turning now to the nonplanar diagrams, there are again two 
\begin{equation*}
-i \Gamma^{(2),s,np}_1 \quad = \quad  \includegraphics[width=0.25\linewidth,valign=c]{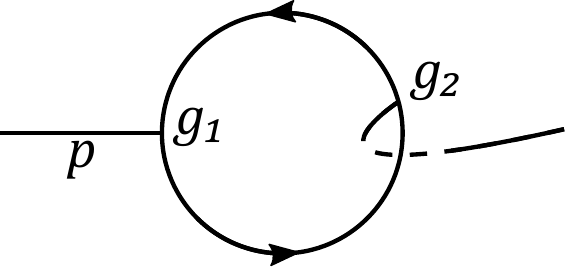} \quad + \quad \includegraphics[width=0.25\linewidth,valign=c]{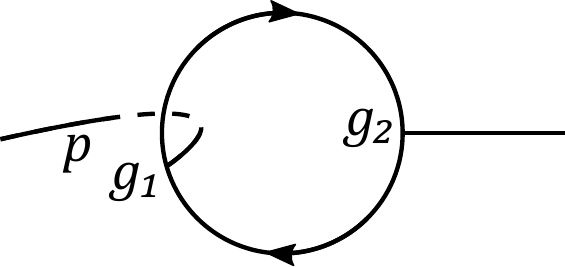}
\end{equation*} 

Each now has one $g_1$ vertex and one $g_2$ vertex, which makes it clear why the analysis of \cite{Anisimov:2001zc} found no such diagrams. The two diagrams will come with opposite phase factors, $e^{i p \wedge k}$ and $e^{i k \wedge p}$, so we can compute one and then find the other by taking $p \mapsto -p$. In this case it's obvious that after completing the square we will only be left with terms which are quadratic in $p$, and so the two diagrams give the same contribution. We can thus compute both terms at the same time.

The phase factor in the integrand will modify our change of variables, as it did in the $\phi^4$ case, to give again an effective cutoff for this diagram due to the noncommutativity. We find
\begin{multline}
\Gamma^{(2),s,np}_1(p) = \frac{g_1 g_2}{\pi^2} \int \text{d}q \text{d}\alpha_1 \text{d}\alpha_2 q^3\left( M^2 - q^2 + \frac{\alpha_1 \alpha_2}{\left(\alpha_1 + \alpha_2\right)^2} p^2+\frac{p \circ p }{4 (\alpha_1 + \alpha_2)^2}\right) \\ \times e^{- \left(\alpha_1 + \alpha_2\right)\left(q^2 + M^2\right) - \frac{\alpha_1\alpha_2}{\alpha_1 + \alpha_2} p^2-\frac{p \circ p }{4 (\alpha_1 + \alpha_2)}}.
\end{multline}
 
We can now follow the same steps to regulate and integrate this, and again find a closed-form expression for the pieces which contain divergences.  Note that unlike the $\phi^4$ calculation, we can already see that the nonplanar expression will not merely be given by $\Lambda \rightarrow \Lambda_{\text{eff}}$, as the change of variables has here modified the numerator of the integrand to give an extra piece to the momentum polynomial multiplying the exponential. And so integration gives us
\begin{multline}  \label{nonplanarscalar}
\Gamma^{(2),s,np}_1(p) = \frac{g_1 g_2}{1920 \pi^2} \left[3\left(640 M^2 + p^4 p \circ p + 40 (4 M^2 + p^2) p \circ p \Lambda_{\text{eff}}^2\right) K_0\left(\frac{\sqrt{4 M^2 + p^2}}{\Lambda_\text{eff}}\right) \right. \\ + \left. 20 \sqrt{4 M^2 + p^2} \Lambda_{\text{eff}} \left(- 96 + p^2 p\circ p + 12 p \circ p \Lambda_{\text{eff}}^2\right) K_1\left(\frac{\sqrt{4 M^2 + p^2}}{\Lambda_\text{eff}}\right)  \right].
\end{multline}

We must now think slightly more carefully about what we want to add to the quadratic effective action to find a Wilsonian interpretation of this theory. We may isolate the IR divergence that appears when the cutoff is removed by first taking the limit $\Lambda \rightarrow \infty$ with $p \circ p$ held fixed, and then expanding around $p \circ p = 0$. We may then ask that this same divergence appears at any value of $\Lambda$. To account for this IR divergence, we must add to our  effective action 
\begin{equation}
\Delta S_{1\text{PI}}(\Lambda) = - \half \int \frac{\text{d}^4p}{(2 \pi)^4} \frac{g_1 g_2}{2 \pi^2} \left(\Lambda_{\text{eff}}^2 - \frac{4}{p\circ p}\right)\varphi(p)\varphi(-p),
\end{equation}
which can easily be done through the addition of an auxiliary scalar field as was done in Section \ref{sec:phi4} and is discussed in more generality in Appendix \ref{app:auxfield}. After having added this to our action, for small $p \circ p$ the scalar two-point function now behaves as $\Gamma_1^s(p) = - \frac{2 g_1 g_2}{\pi^2 p \circ p} + \dots$ for any value of $\Lambda$. The new pole in this case has the opposite sign as that in \ref{eqn:auxaction}, and so will be accessible in Euclidean signature, clearly signaling a tachyonic instability. While this puts the violation of unitarity in this theory on prime display, it also means that this pole will be accessible in the $s$-channel in the Lorentzian theory if we allow for timelike noncommutativity. 

We emphasize that any conclusions about the Lorentzian theory with timelike noncommutativity are speculative and dependent upon a solid theoretical understanding of a unitary formulation of the field theory, and in principle such a formulation could find radically different IR effects than this na\"{i}ve approach. However, it was found in \cite{Bozkaya:2002at} that a modification of time-ordering to explicitly make the theory unitary (at the expense of microcausality violation) leaves the one-loop correction to the self-energy unchanged in $\phi^4$ theory, and the same might be expected to hold true for Yukawa theory. This makes it worthwhile to at least briefly consider the potential phenomenological consequences of the new pole.

At low energies, the propagator is here modified to $m^2 + (p_i + p_j)^2 - \frac{2 g_1 g_2}{\pi^2} \frac{1}{(p_i + p_j)\circ (p_i + p_j)}$. If we consider scattering of fermions through an $s$-channel $\varphi$ and take the simple case of a noncommutativity tensor which in the lab frame has one eigenvalue $1/\Lambda_\theta^2$ with $m^2 \gg \Lambda_\theta^2$, then the emergent pole appears at $s = \frac{2 g_1 g_2}{\pi^2} \frac{1 - \beta^2}{1+\beta^2} \frac{\Lambda^4_\theta}{m^2}$. Here $s = - (p_i + p_j)^2$ is the invariant momentum routed through the propagator, and $\beta$ is the boost of the $(p_i + p_j)$ system with respect to the lab frame. The Lorentz-violation here then has the novel effect of smearing out the resonance corresponding to the light pole for a particle which is produced at a variety of boosts. This is in contrast to the pole at $m^2$, which gives a conventional resonance at leading order. Of course, we have not constructed a fully realistic theory in any respect, and ultimately it may well be that other Lorentz-violating effects provide the leading constraint. Nonetheless, the lineshape of resonances may be an interesting observable in this framework.

A further feature of this opposite sign of the new pole compared to that in the $\phi^4$ theory is that the unusual momentum-dependence of the two-point function will lead to ordered phases which break translational invariance \cite{Gubser:2000cd,Minwalla:1999px,Steinacker:2005wj,Chen:2001an,Castorina:2003zv}. 
While a Lorentz-violating background field may possibly be very well constrained, the detailed constraint depends on its wavelength and the ways in which it interacts with the SM. But this is another obvious line of exploration for constraining realistic NCFTs.

\subsection{Fermion Two-Point Function}
 
There are again two planar diagrams:
\begin{equation*}
-i\Gamma^{(2),f,p}_1 \quad = \quad  \includegraphics[width=0.25\linewidth,valign=c]{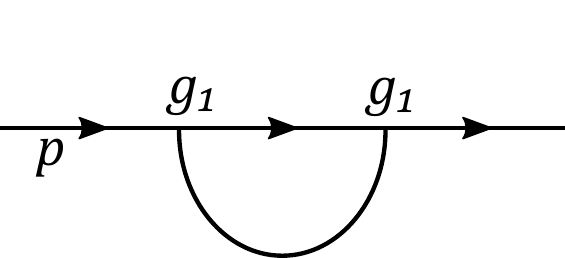} \quad + \quad \includegraphics[width=0.25\linewidth,valign=c]{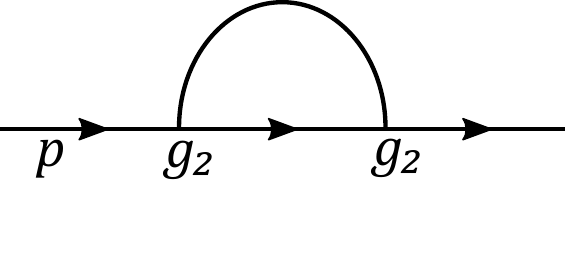}
\end{equation*} 

\noindent No new features appear in the evaluation of these diagrams, so we merely quote the final result:
\begin{equation}
\Gamma^{(2),f,p}_1 = -\frac{g_1^2 + g_2^2}{16 \pi^2} \left(M - \frac{\slashed{p}}{2}\right) \log\frac{4 p^2 \Lambda^2}{m^4 + 2 m^2 (p^2 - M^2) + (M^2 + p^2)^2} + \dots 
\end{equation}

\noindent We also have two nonplanar diagrams, which again mix the two vertices
\begin{equation*}
-i\Gamma^{(2),f,np}_1 \quad = \quad  \includegraphics[width=0.25\linewidth,valign=c]{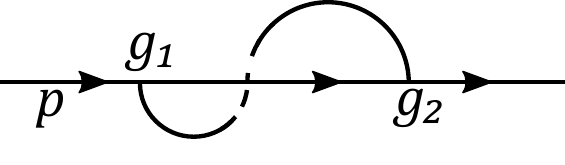} \quad + \quad \includegraphics[width=0.25\linewidth,valign=c]{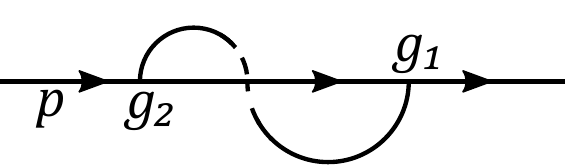}
\end{equation*} 

Here we find that the different phase factors for each diagram, which we saw were inconsequential for the nonplanar corrections to the scalar, have an important role. When we complete the square in each of the two cases, we find that one of the diagrams has an integrand proportional to $\left(M - \slashed{p} \frac{\alpha_2}{\alpha_1 + \alpha_2} - \frac{1}{2}\frac{p_\mu\theta^{\mu\nu}\gamma_\nu}{\alpha_1 + \alpha_2}\right)$ and the other is proportional to $\left(M - \slashed{p} \frac{\alpha_2}{\alpha_1 + \alpha_2} + \frac{1}{2}\frac{p_\mu\theta^{\mu\nu}\gamma_\nu}{\alpha_1 + \alpha_2}\right)$, so the would-be divergence in $p\theta$ will cancel manifestly between the two diagrams. After this everything proceeds as before, and we find 
\begin{equation}
\Gamma^{(2),f,np}_1 = -\frac{g_1 g_2}{8 \pi^2} \left(M - \frac{\slashed{p}}{2}\right) \log\frac{4 p^2 \Lambda_{\text{eff}}^2}{m^4 + 2 m^2 (p^2 - M^2) + (M^2 + p^2)^2} + \dots
\end{equation}

We see that with $g_1 = g_2 \equiv g$, the fermion quadratic effective action also behaves as expected from `strong UV/IR duality'. The logarithmic divergence of the commutative theory has been transmogrified in the nonplanar diagrams into IR dynamics via the simple replacement $\Lambda \rightarrow \Lambda_{\text{eff}}$, and so a $p \circ p \rightarrow 0$ pole will emerge when we remove the cutoff. We discuss the use of an auxiliary field to restore a Wilsonian interpretation here in Appendix \ref{app:auxferm}.

\subsection{Three-Point Function}

The correction to the vertex function constitutes further theoretical data toward the Wilsonian interpretation of the noncommutative corrections. We calculate the one-loop correction in this section and delay the discussion of the use of auxiliary fields to account for them until Appendix \ref{app:aux3pt}. We will find that while we can use the same fields to account for the modifications to both the propagators and the vertices, the physical interpretation of such fields is unclear.

We can compute corrections for each fixed ordering of external lines separately since they're coming from different operators. For simplicity we'll compute the $g_1$ ordering, which we will denote $\Gamma^{\varphi\overline{\psi}\psi}_3(r,p,\ell)$. There are four diagrams in total: one planar diagram with two insertions of the $g_2$ vertex, one nonplanar diagram with two insertions of the $g_1$ vertex, and two nonplanar diagrams with one insertion of each. It is easy to see by looking at the diagrams that the same expressions with $g_1 \leftrightarrow g_2$ compute the correction to the other ordering, $\Gamma_3^{\overline{\psi}\varphi\psi}(r,p,l)$.

The new feature of this computation is that we now need \textit{three} Schwinger parameters, and this presents a problem for our previous computational approach. We won't be able to perform the two finite integrals before expanding in a variable which isolates the divergences when $\alpha_1 + \alpha_2 + \alpha_3 \rightarrow 0$, analogously to what we did in $2d$ Schwinger space. Instead we slice $3d$ Schwinger space such that we can perform the integral which isolates the leading divergences first, and then --- as long as we're content only to understand this divergence --- we can discard the rest without having to worry about performing the other two integrals.

The planar diagram is 
\begin{equation*}
i\Gamma^{\varphi\overline{\psi}\psi}_{3,p}(p,\ell) \quad = \quad  \includegraphics[width=0.25\linewidth,valign=c]{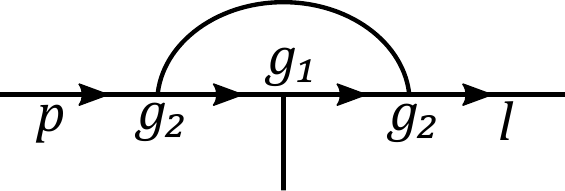},
\end{equation*} 

and corresponds to the expression
\begin{equation}
\Gamma^{\varphi\overline{\psi}\psi}_{3,p}(p,\ell) = -i (i g_1) (i g_2)^2 \int \frac{\text{d}^4k}{(2\pi)^4} \frac{(-i)^3 \left(M - (\slashed{k} + \frac{\slashed{p}}{2} + \frac{\slashed{\ell}}{2})\right)\left(M - (\slashed{k}- \frac{\slashed{p}}{2} - \frac{\slashed{\ell}}{2}\right)}{\left((k+\frac{p}{2}+\frac{\ell}{2})^2 + M^2\right)\left((k-\frac{p}{2}-\frac{\ell}{2})^2 + M^2\right)\left((k+\frac{p}{2}-\frac{\ell}{2})^2 + m^2\right)}.
\end{equation}
After moving to Schwinger space, integrating over the loop momentum, and introducing a cutoff $\exp\left(-1/\left(\Lambda^2 (\alpha_1 + \alpha_2 + \alpha_3)\right)\right)$, we switch variables to 
\begin{equation}
\alpha_1 = \xi_1 \eta, \qquad 
\alpha_2 = \xi_2 \eta, \qquad
\alpha_3 = (1 - \xi_1 - \xi_2) \eta,
\end{equation}
under which $\int_0^\infty \text{d}\alpha_1 \int_0^\infty \text{d}\alpha_2 \int_0^\infty \text{d}\alpha_3 \rightarrow \int_0^1 \text{d} \xi_1 \int_0^{1-\xi_1} \text{d} \xi_2 \int_0^\infty \text{d} \eta \ \eta^2$. Performing the momentum integral transfers the divergence for large $k$ to a divergence in small $\alpha_1 + \alpha_2 + \alpha_3 = \eta$. This will allow us to find the leading divergent behavior immediately by carrying out the $\eta$ integral and then expanding in $\Lambda \rightarrow \infty$. This yields
\begin{equation}
\Gamma^{\varphi\overline{\psi}\psi}_{3,p}(p,\ell) = \frac{g_1 g_2^2}{16 \pi^2} \log\left(\Lambda^2\right) + \text{ finite},
\end{equation}
where we are unable to determine the IR cutoff of the logarithm, but this suffices for our purposes.

The three nonplanar graphs now each receive a different phase corresponding to which external line crosses the internal line
\begin{equation*}
i\Gamma^{\varphi\overline{\psi}\psi}_{3,np}(p,\ell) \quad = \quad  \includegraphics[width=0.25\linewidth,valign=c]{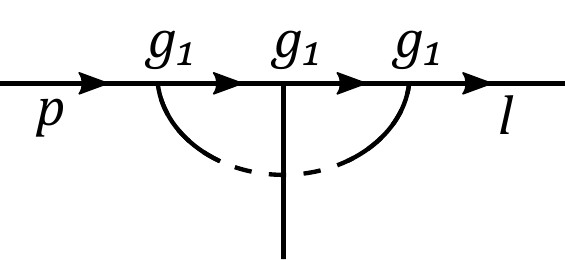} \quad + \quad \includegraphics[width=0.25\linewidth,valign=c]{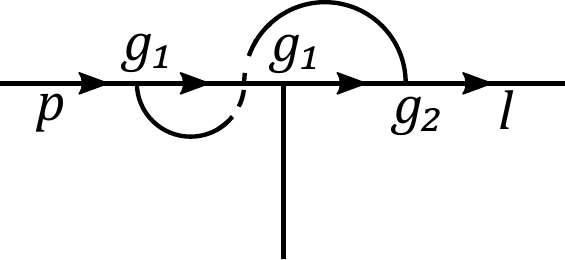} \quad + \quad \includegraphics[width=0.25\linewidth,valign=c]{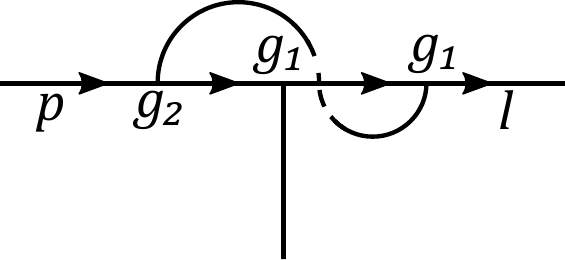},
\end{equation*} 
where the first gets $\exp\left[-i(k \wedge p + k \wedge \ell + p \wedge \ell)\right]$, the second $\exp\left[-i(k \wedge p + p \wedge \ell/2)\right]$, and the third $\exp\left[-i(k \wedge \ell + p \wedge \ell/2)\right]$. The evaluation of these diagrams proceeds as in the previous examples. If we take the IR limit $p,\ell \rightarrow 0$ of the nonplanar contributions to this ordering of the three-point function and then expand in large $\Lambda$ we find 
\begin{equation}
\lim\limits_{p,\ell\rightarrow 0} \Gamma^{\varphi\overline{\psi}\psi}_{3,np}(p,\ell) = \frac{g_1^2 (g_1 + 2 g_2)}{16 \pi^2} \log\left(\Lambda^2\right) + \text{ finite}.
\end{equation}
However, if we first take the UV limit $\Lambda \rightarrow 0$, and then expand in small momenta, we find
\begin{equation}
\lim\limits_{\Lambda\rightarrow \infty} \Gamma^{\varphi\overline{\psi}\psi}_{3,np}(p,\ell) = \frac{g_1^2}{16 \pi^2} \left[g_1 \log\left(\frac{4}{(p + \ell) \circ (p + \ell)}\right) + g_2 \log\left(\frac{4}{p \circ p}\right)+ g_2 \log\left(\frac{4}{\ell \circ \ell}\right)\right] + \text{ finite}, \label{eqn:3ptfct}
\end{equation}
where we again see UV/IR mixing, and we note that each nonplanar diagram has been effectively cutoff by the momenta which cross the internal line. We discuss the use of auxiliary fields to restore a Wilsonian interpretation to this vertex correction in Appendix \ref{app:aux3pt}.

%% file: wesszumino_v5.tex
We now turn our attention to the softly-broken noncommutative Wess-Zumino model as a controllable example of the interplay between UV/IR mixing and the finiteness of the field theory. We will restrict ourselves to calculating the one-loop correction to the scalar two-point function. Since the new poles appearing in the quadratic effective action in the scalar and Yukawa theories are intimately related to the quadratic divergences of the commutative theories, we will not be surprised to find that this feature will disappear when both the scalar and the fermion are present in the EFT below the cutoff. By studying the softly-broken theory we can take the fermion above or below the cutoff to smoothly see the relation between the finiteness of the field theory and the effects of UV/IR mixing. The exactly supersymmetric noncommutative Wess-Zumino model was first discussed in detail in \cite{Girotti:2000gc}, and the absence of an infrared pole in a softly-broken theory was first noted in \cite{Matusis:2000jf}. The softly-broken Wess-Zumino model was first considered in \cite{AmelinoCamelia:2002au}.\footnote{Our one-loop results agree with those of \cite{AmelinoCamelia:2002au} save for their claim that logarithmic IR divergences are absent in the exactly supersymmetric theory, which contradicts \cite{Girotti:2000gc}. We will below find a logarithmic IR divergence in the wavefunction renormalization which is independent of the soft-breaking, which is consistent with the expectations of strong UV/IR duality.}

The noncommutative Wess-Zumino theory can be suitably formulated in off-shell superspace as
\begin{equation}
\mathcal{L} = \int \text{d}^4\theta \ Z \Phi^\dagger \Phi + \int \text{d}^2\theta \ \left(\half M \Phi^2 + \frac{1}{6} y \ \Phi \star \Phi \star \Phi\right) + \text{ h.c.},
\end{equation}
where $\Phi$ is a chiral superfield and we have included a wavefunction renormalization factor in the K\"{a}hler potential $Z = 1 + \mathcal{O}(y^2)$. We can introduce soft supersymmetry breaking by promoting this factor to a spurion $Z = 1 + ( \left|M\right|^2 - m^2)  \theta^2 \theta^{\dagger2}$, the only effect of which is to modify the scalar mass spectrum.

Formulating the noncommutative theory including the auxiliary $F$ fields makes it manifest that we have preserved supersymmetry off-shell. This procedure is in fact precisely the same as quantizing after integrating out $F$, and so we end up with a star-product version of the familiar Lagrangian:
\begin{align}
-\mathcal{L}_{\text{NCWZ}} &= Z \partial^\mu \phi^{*} \partial_\mu \phi - i Z \psi^\dagger \bar{\sigma}^\mu \partial_\mu \psi \nonumber\\ 
&+ Z^{-1} m^2 \phi^{*} \phi + \half M \psi \psi + \half M^{*} \psi^\dagger \psi^\dagger \nonumber\\ 
&+ \half Z^{-1} y \phi \star \psi \star \psi + \half Z^{-1} y^{*} \phi^{*} \star \psi^\dagger \star \psi^\dagger  \nonumber\\
&+ \half Z^{-1} y M^{*} \phi \star \phi \star \phi^{*} + \half  Z^{-1} y^{*} M \phi^{*} \star \phi^{*} \star \phi \nonumber\\
&+ \frac{1}{4} Z^{-1} \left|y\right|^2 \phi \star \phi \star \phi^{*} \star \phi^{*}
\end{align}

\noindent where $\phi$ is a complex scalar and $\psi$ is a Weyl fermion. Of course, now that we've introduced supersymmetry breaking we expect to find that there is further renormalization beyond that associated with $Z$, but keeping the manifest factors of $Z$ will allow us to easily compare to our expectations for the supersymmetric limit.

The calculation of the one-loop correction to the two-point function goes much as the previously-demonstrated examples. The presence of the three-scalar interaction gives a new class of diagrams, whose evaluation is routine. The two-component fermions yield slightly different factors than did the Dirac fermions \cite{Dreiner:2008tw}. Finally, it is important to note that the results for the diagrams computed in Section \ref{sec:phi4} cannot be used here, as we must here regulate uniformly using $\exp(-1/(\Lambda^2 (\alpha_1 + \alpha_2)))$ like we did in Section \ref{sec:yukawa}. This may be easily accommodated by writing the integrand in the quartic diagrams as $\frac{1}{k^2 + m^2} \frac{k^2 + m^2}{k^2 + m^2}$.

Adding up all these diagrams and taking the limit where $\Lambda, \Lambda_{\text{eff}}$ are large, we find that the one-loop scalar two-point function may be organized as 
\begin{align}
\Gamma^{(2),s} &\equiv Z p^2 + Z^{-1}(m^2 + \delta m^2) \\
Z &= 1 + \frac{y^2}{32 \pi^2} \log\left[\frac{\Lambda \Lambda_{\text{eff}}}{M^2}\right] + \dots \\
\delta m^2 &= \frac{y^2}{32 \pi^2} \left(M^2 -  m^2\right) \log\left[ \frac{\Lambda \Lambda_{\text{eff}}}{M^2}\right] + \dots,
\end{align}

\noindent where we make manifest the presence of supersymmetric nonrenormalization in the limit $m \rightarrow M$, which acts as a non-trivial check. As expected, the absence of the quadratic UV divergence in the Wess-Zumino model has led to the absence of an infrared pole from the noncommutativity, even as the fermion is made arbitrarily heavy relative to the scalar. However, logarithmic UV/IR mixing still occurs.  

We may repeat this calculation using dimensional regularization and taking note of the issues which arose in Section \ref{sec:dimreg}. Using the same parametrization of the one-loop two-point function as above, the planar diagrams contribute
\begin{align}
Z_{\text{planar}} &= 1 + \frac{y^2}{64 \pi^2} \left(\frac{2}{\epsilon} + \log \frac{\mu^2}{M^2}\right) + \dots \\
\delta m^2_{\text{planar}} &= \frac{y^2}{64 \pi^2} (M^2 - m^2) \left(\frac{2}{\epsilon} + \log \frac{\mu^2}{M^2}\right) + \dots,
\end{align}
as expected. The full form of the nonplanar diagrams is unenlightening, but if we take the IR limit $p \circ p \rightarrow 0$ first, they give precisely the same contribution as the planar diagrams, since the diagram degeneracies are all the same in this case. Taking the UV limit $\epsilon \rightarrow 0$ first (and staying in $d < 2$), we instead find 
\begin{align}
Z_{\text{nonplanar}} &= 1 + \frac{y^2}{64 \pi^2} \log \frac{4}{M^2 p \circ p} + \dots \\
\delta m^2_{\text{nonplanar}} &= \frac{y^2}{64 \pi^2} (M^2 - m^2) \log \frac{4}{M^2 p \circ p} + \dots,
\end{align}
which has precisely the same correspondence with the Schwinger-space regularization as we saw for the $\phi^4$ case.

We thus see clearly the conflict between supersymmetry and the use of UV/IR mixing to explain low-energy puzzles. UV/IR mixing transmogrified UV momentum dependence into IR momentum dependence, and so depended crucially on the sensitivity of our field theory to UV modes. For a theory which is finite as a field theory, the dependence on the UV physics has been removed, and so we see no interesting IR effects. 

Of course, in the presence of a cutoff $\Lambda$ it is also possible to study the behavior of the scalar two-point function when $M^2 \gg \Lambda^2 \gg |M^2-m^2|$ as the fermion is taken above the cutoff while keeping the scalar light. This corresponds to taking $M/\Lambda,M/\Lambda_{\text{eff}} >1$ and then expanding in the limit where $\Lambda, \Lambda_{\text{eff}}$ are large. This gets rid of the nonplanar Yukawa-type diagrams and, as one might expect, results in a return of UV sensitivity in the scalar EFT below the cutoff, foreshadowing a return of the UV/IR mixing effects. The scalar mass-squared in this limit becomes
\begin{equation}
\delta m^2 = \frac{y^2}{256 \pi^2} \left(6 M^2 + 16 \Lambda^2 + 8 \Lambda_{\text{eff}}^2 \right) + \dots.
\end{equation}  
and UV/IR mixing reappears at the quadratic level. So our EFT intuition isn't totally out the window; it's been broken in a controlled way, and we can smoothly interpolate between theories with and without UV/IR mixing by taking the states responsible for finiteness above the cutoff. This sharpens the sense in which UV/IR mixing can do something interesting in the IR as long as the field-theoretic description of our universe is never finite. 

Ultimately, this highlights a central challenge for approaching the hierarchy problem via UV/IR mixing. The hierarchy problem is particularly sharp when the full theory is finite and scale separation is large, in which case the sensitivity of the Higgs mass to underlying scales is unambiguous. But UV/IR mixing effects potentially relevant to the hierarchy problem are absent in this case, and emerge only when finiteness is lost. This tension is not necessarily fatal to UV/IR approaches to the hierarchy problem -- ultimately the UV sensitive degrees of freedom are not the ones we would wish to identify with the Higgs -- but it bears emphasizing.

Moreover, there is a possible loophole in the general argument that finiteness must be surrendered in order to generate a scale from UV/IR mixing. The presence of interesting effects in the IR here depends solely on the UV sensitivity of the nonplanar diagrams. The `orbifold correspondence' \cite{Kachru:1998ys,Bershadsky:1998cb,Schmaltz:1998bg} provides non-supersymmetric field theories constructed via orbifold truncation of $\mathcal{N}>0$ theories whose planar diagrams agree with those of the supersymmetric theory and so are finite. A noncommutative orbifold field theory \cite{Armoni:1999nj} may then provide a theory which is fully predictive, yet which still generates an infrared scale via UV/IR mixing. Generally, it may be possible that UV/IR mixing appears in such a way that it is the sole effect sensitive to short distances.

%% file: lessons_v5.tex
To attempt to formulate a realistic theory which uses UV/IR mixing to solve extant theoretical puzzles, it would be useful to have an understanding of which features of NCFT were responsible for the curious infrared effects discussed above. This would be helpful whether one wishes to test out these ideas in any of the many proposed modifications of NCFT, or to write down other toy models which share some features of NCFT but are based upon different principles.

Qualitatively, the two unusual features involved in the formulation of NCFT are Lorentz invariance violation and nonlocality. However, it is obvious that one may have theories with one or both of these features without the interesting effects we have seen. The answer then is not so simple as pointing to one axiom or another of EFT which has been broken, but depends sensitively on the way in which they are broken. We briefly explore two ways we may better understand the interplay here between nonlocality and Lorentz-violation and how they come together to cause surprising low-energy effects. We first give a general argument based on the way nonlocality appears to postdict the form of the violation of EFT expectations. We then phenomenologically examine the loop integration appearing in our NCFT calculations to diagnose what caused the appearance of the IR pole. This will lead us to discuss an avenue toward investigating (or manufacturing) such effects in nonlocal, Lorentz-invariant theories.

To see how EFT expectations may be violated, consider the peculiar way in which the noncommutative effects in the one-loop action (e.g. Equation \ref{eqn:1PIaction}) induce nonlocality. In Wilsonian EFT, integrating out momentum modes $p \gtrsim \Lambda$ produces a nonlocal theory at those scales, or equivalently on distances $x \lesssim 1/\Lambda$. However, particles on a noncommutative space can be thought of as rods of size $L \sim p \theta$ \cite{SheikhJabbari:1999vm,Bigatti:1999iz,Seiberg:2000gc,Girotti:2001dh,Acatrinei:2002sb}. This tells us that in a NCFT we should expect nonlocality to be present for scales $x \lesssim p \theta$. Comparing the two scales, we see that we should find nonlocal effects past those expected in Wilsonian EFT for $\frac{1}{\Lambda} < p\theta$. Here this momentum-dependent nonlocality occurs in a Lorentz-violating way. This expectation was exactly borne out in the examples above, where we saw that the one-loop effective action in momentum space is nonlocal for $p\circ p \gg 1/\Lambda^2$ \cite{Minwalla:1999px}. 

Purely from this analysis of the form of nonlocality, we may conclude there will be a breakdown of Wilsonian renormalization. After we remove the cutoff, the theory should be nonlocal on all scales $p \circ p > 0$. But if we compute a correlation function at a large-but-finite $\Lambda$, the theory will still be local for momenta $p \circ p < 1/\Lambda^2$, and so will greatly differ from the continuum result. So our surprising discovery of the non-uniform convergence of correlation functions in the examples above is understood easily from this picture. 

While this sort of momentum-dependent nonlocality may seem \textit{ad hoc}, it has been suggested previously for separate purposes. It has been argued \cite{Maggiore:1993rv} that quantum gravity should obey a `Generalized Uncertainty Principle' $\Delta x \gtrsim \frac{\hbar}{\Delta p} + \ell_p^2 \Delta p$, with $\ell_p$ the Planck length, based on the use of Hawking radiation to measure the horizon area of a black hole. This gives precisely the same sort of momentum-dependent nonlocality as we saw above. We refer the reader to \cite{Tawfik:2015rva} for a review of the Generalized Uncertainty Principle, \cite{Konishi:1989wk,Yoneya:2000bt} for similar conclusions within string theory, and \cite{Hossenfelder:2012jw} for a more general review of the appearance of an effective minimal length in quantum gravity. It would be interesting to investigate other field theories which obey such uncertainty principles and determine whether UV/IR mixing causes similar features as appear in NCFT. For theories which violate Lorentz invariance, care must be taken to avoid arguments that even Planck-scale Lorentz violation is empirically ruled out \cite{Collins:2004bp,Polchinski:2011za}.

We may also attempt to phenomenologically diagnose what caused the appearance of the IR pole from the form of the loop integration. The presence of an exponential of momenta was clearly crucial, and this implies a necessity of nonlocality. It's also clear that the modification of the cutoff in the nonplanar diagrams $\Lambda \mapsto \Lambda_{\text{eff}}$, which rendered the diagrams UV finite in a way that brought UV/IR mixing, was a result of the contraction between the loop momentum and the external momentum. Less obviously, one may see that any quadratic term in loop momentum in the exponential would have erased this feature, as after momentum integration one would find an integrand $\sim \frac{1}{1 + \alpha_+}$, and any divergence will have disappeared. Heuristically, the quadratic suppression in loop momentum is too strong and regulates the UV divergence entirely independently of the cutoff, so no UV/IR mixing appears. NCFT disallows such terms as a result of momentum contractions being performed with an antisymmetric tensor, and this particular mechanism seems to imply the necessity of Lorentz invariance violation. However, this argument only considers small deviations from the form of the integral in NCFT. Further discussions of the form of loop integrals with generalizations of the star-product may be found in \cite{Galluccio:2009ss,Ardalan:2010ht}.

Likely a better approach to understand the prospect for finding features similar to that of NCFT in a Lorentz invariant theory is to back up and study formulations of Lorentz invariant extensions of NCFT. This is accomplished by upgrading the noncommutativity tensor $\theta^{\mu\nu}$ from a $c$-number to an operator. This was proposed already by Snyder in 1947 \cite{Snyder:1946qz}, and this approach has been revived a number of times more recently (e.g.  \cite{Doplicher:1994tu,Kase:2002pi,Carlson:2002wj,Heckman:2014xha,Much:2017hcv}). Schematically, this results in an action containing an integral over $\theta^{\mu\nu}$ 
\begin{equation}
S = \int \text{d}^4x \ \text{d}^6\theta \ W(\theta) \ \mathcal{L}(\phi, \partial \phi),
\end{equation}
where $W(\theta)$ is a `weighting function', and the Lagrangian is still defined using the star-product. The challenge in this approach for our purposes is in devising a method for nonperturbative calculations in $\theta$, which as we saw above was necessary to preserve the features of UV/IR mixing.

Searching more generally for Lorentz invariant theories which contain UV/IR mixing will likely allow more promising phenomenological applications. That such theories should exist can be broadly motivated by quantum gravity, as any gravitational theory is expected both to be nonlocal and to have UV/IR mixing. That Lorentz violation should be present is less clear. A particularly interesting line of development is to then understand in detail the class of nonlocal theories that would have UV/IR mixing of a sort similar to that discussed here. Recent work toward placing nonlocal quantum field theories on solid theoretical ground \cite{Tomboulis:2015gfa,Chin:2018puw} is clearly of sharp interest here, though the larger goal is quite distinct. The nonlocality studied in these works is designed to render the field theory UV-finite, and so the nonlocal vertex kernels are chosen precisely to avoid the introduction of new poles by ensuring these are momentum-space entire functions which vanish rapidly in Euclidean directions. The nonlocal vertices of NCFT manage to introduce new poles by oscillating as $p \rightarrow \infty$, which presumably allows for the appearance of new `endpoint singularities' \cite{Eden:1966dnq,Itzykson:1980rh}, though a full examination of the Landau equations in NCFT has not (to our knowledge) been performed. Our interest is thus in a disjoint class of nonlocal theories, where new poles can appear in interesting ways. Classifying the space of such theories and developing an approach to systematically understand their unitarity properties seems well motivated.

%% file: conclusions_v5.tex
The lack of evidence for conventional solutions to the hierarchy problem has placed particle physics at a crossroads. While it is possible that the answer ultimately lies further down the well-trodden path of existing paradigms, the appeal of less-travelled paths grows greater with every inverse femtobarn of LHC data. 

In this work we have ventured to take seriously the apparent failure of expectations from Wilsonian effective field theory regarding the hierarchy problem by investigating a concrete framework --- noncommutative field theory --- in which Wilsonian EFT itself breaks down. Not only does noncommutative field theory violate Wilsonian expectations, it provides a sharp instance of UV/IR mixing: ultraviolet modes of noncommutative theories can generate an infrared scale whose origin is opaque to effective field theory. To the extent that UV/IR mixing has any relevance to the hierarchy problem, the emergence of an infrared scale seems to be among the most promising effects. Although the real-world applicability of these theories is likely limited by their Lorentz violation, they nonetheless provide valuable toy models for exploring the potential relevance of UV/IR mixing to problems of the Standard Model. 

To this end, we have surveyed existing results on noncommutative theories with an eye towards `strong UV/IR duality' --- the transmogrification of UV divergences into infrared poles at the same order. This led us to a detailed analysis of noncommutative Yukawa theory, perhaps the most useful toy model for thinking about the hierarchy problem (insofar as the Yukawa sector of the Standard Model is responsible for the largest UV sensitivity of the Higgs mass, and highlights the relative UV {\it insensitivity} of the fermion masses). In the noncommutative theory, the presence of both inequivalent Yukawa couplings implies the same strong UV/IR duality exhibited by real $\phi^4$ theory: a quadratic divergence in the one-loop correction to the scalar mass from fermion loops gives rise to a simple IR pole, while a logarithmic UV divergence in the one-loop correction to the fermion mass from scalar loops give rise to only a logarithmic IR divergence. Intriguingly, the infrared pole in the scalar two-point function appears accessible in the $s$-channel in the Lorentzian theory, a feature which gives it particular phenomenological relevance. 

We then introduced softly-broken supersymmetry as a way to explore the interplay between (in)finiteness and UV/IR mixing. Choosing soft terms in order to keep the scalar light as the fermion mass is varied concretely illustrates several expected features. Strong UV/IR duality is preserved in the sense that both UV and IR divergences are absent at quadratic order (and persist at logarithmic order) when both the scalar and the fermion are in the spectrum. However, infrared structure reappears as the fermion mass is raised above a fixed cutoff and (quadratic) finiteness is lost. This underlines the sense in which UV/IR mixing may only ever play an interesting role when the field theory is quadratically UV sensitive at all scales, a scenario in which the hierarchy problem is less concrete.

Finally, building on the lessons from the toy models considered here, we have highlighted a variety of interesting lines of exploration in theories featuring nonlocality with or without Lorentz violation that may be of relevance to the hierarchy problem.

While the prospect that UV/IR mixing will solve outstanding theoretical problems in the low-energy universe is possibly fanciful, now is the time for such reveries. The paradigms of the past few decades of particle theory are under considerable empirical pressure, and innovative approaches are needed. At the very least, by pushing the limits of EFT we stand to learn more about the broad spectrum of phenomena possible within quantum field theory.

%% file: auxfield_v5.tex
In this appendix we discuss various generalizations of the procedure introduced in \cite{Minwalla:1999px,VanRaamsdonk:2000rr} to account for the new structures appearing in the noncommutative quantum effective action via the introduction of additional auxiliary fields. 

\subsection{Scalar Two-Point Function}

It is simple to generalize the procedure discussed in Section \ref{sec:phi4} to add to the quadratic effective action of $\phi$ any function we wish through judicious choice of the two-point function for an auxiliary field $\sigma$ which linearly mixes with it. In position space, if we wish to add to our effective Lagrangian 
\begin{equation}
\Delta \mathcal{L}_{\text{eff}} = \half c^2 \phi(x) f(-i\partial) \phi(x),
\end{equation} 
where $f(-i\partial)$ is any function of momenta, and $c$ is a coupling we've taken out for convenience, then we simply add to our tree-level Lagrangian 
\begin{equation}
\Delta \mathcal{L} = \half \sigma(x) f^{-1}(-i \partial) \sigma(x) + i c \sigma(x) \phi(x),
\end{equation}
where $f^{-1}$ is the operator inverse of $f$.  It should be obvious that this procedure is entirely general. As applied to the Euclidean $\phi^4$ model, we may use this procedure to add a second auxiliary field to account for the logarithmic term in the quadratic effective action as
\begin{equation}
\Delta \mathcal{L} = \half \sigma(x) \frac{1}{\log\left[1 - \frac{4}{\Lambda^2 \partial \circ \partial }\right]} \sigma(x) - \frac{g M}{\sqrt{96 \pi^2}} \sigma(x) \phi(x),
\end{equation}
where we point out that the argument of the log is just $4/(\Lambda_{\text{eff}}^2 p \circ p)$ in position space. We may then try to interpret $\sigma$ also as a new particle. As discussed in \cite{VanRaamsdonk:2000rr}, its logarithmic propagator may be interpreted as propagation in an additional dimension of spacetime.

Alternatively, we may simply add a single auxiliary field which accounts for both the quadratic and logarithmic IR singularities by formally applying the above procedure. But having assigned them an exotic propagator, it then becomes all the more difficult to interpret such particles as quanta of elementary fields.

\subsection{Fermion Two-Point Function} \label{app:auxferm}

To account for the IR structure in the fermion two-point function, we must add an auxiliary fermion $\xi$. If we wish to find a contribution to our effective Lagrangian of 
\begin{equation}
\Delta \mathcal{L}_{\text{eff}} = c^2 \bar \psi \mathcal{O} \psi,
\end{equation}
where $\mathcal{O}$ is any operator on Dirac fields, then we should add to our tree-level Lagrangian
\begin{equation}
\Delta \mathcal{L} = - \bar \xi \mathcal{O}^{-1} \xi + c \left( \bar \xi \psi + \bar \psi \xi\right),
\end{equation}
with $\mathcal{O}^{-1}$ the operator inverse of $\mathcal{O}$. In the Lorentzian Yukawa theory of Section \ref{sec:yukawa}, if we add to the Lagrangian
\begin{equation}
\Delta \mathcal{L} =  - \overline{\xi} \frac{M - i \slashed{\partial}/2}{M^2 - \partial^2/4} \left[\log\left(1 - \frac{4}{\Lambda^2 \partial \circ \partial}\right)\right]^{-1} \xi + \frac{g}{2\sqrt{2}\pi} \left( \overline{\xi} \psi + \overline{\psi} \xi\right).
\end{equation}
we again find a one-loop quadratic effective Lagrangian which is equal to the $\Lambda \rightarrow \infty$ value of the original, but now for any value of $\Lambda$.

\subsection{Three-Point Function} \label{app:aux3pt}

We may further generalize the procedure for introducing auxiliary fields to account for IR poles to the case of poles in the three-point effective action. It's clear from the form of the IR divergences in Equation \ref{eqn:3ptfct} that they `belong' to each leg, and so na\"{i}vely one might think this means that the divergences we've already found in the two point functions already fix them. However those corrections only appear in the internal lines and were already proportional to $g^2$, and so they will be higher order corrections. Instead we must generate a correction to the vertex function itself which only corrects one of the legs. 

To do this we must introduce auxiliary fields connecting each possible partition of the interaction operator. However, while an auxiliary scalar $\chi$ coupled as $\chi \varphi + \chi \overline{\psi}\psi$ would generate a contribution to the vertex which includes the $\chi$ propagator with the $\varphi$ momentum flowing through it, it would also generate a new $(\overline{\psi}\psi)^2$ contact operator, which we don't want. To avoid this we introduce \emph{two} auxiliary fields with off-diagonal two-point functions, a trick used for similar purposes in \cite{VanRaamsdonk:2000rr}. By abandoning minimality, we can essentially use an auxiliary sector to surgically introduce insertions of functions of momenta wherever we want them.

We can first see how this works on the scalar leg. We add to our tree-level Lagrangian 
\begin{equation}
\Delta \mathcal{L} = - \chi_1(x) f^{-1}(-i\partial) \chi_2(x) + \kappa_1 \chi_1(x) \varphi(x) + \kappa_2 \chi_2(x) \star \overline{\psi}(x) \star \psi(x).
\end{equation}
Now to integrate out the auxiliary fields we note that for a three point vertex, one may use momentum conservation to put all the noncommutativity between two of the fields. That is, $\chi_2(x) \star \overline{\psi}(x) \star \psi(x) = \chi_2(x)  (\overline{\psi}(x) \star \psi(x)) =  (\overline{\psi}(x) \star \psi(x)) \chi_2(x)$ as long as this is not being multiplied by any other functions of $x$. So we may use this form of the interaction to simply integrate out the auxiliary fields. We end up with 
\begin{equation}
\Delta \mathcal{L}_{\text{eff}} = \kappa_1 \kappa_2 \overline{\psi} \star \psi \star f(-i\partial) \varphi 
\end{equation}
which is exactly of the right form to account for an IR divergence in the three-point function which only depends on the $\varphi$ momentum.

For the fermionic legs, we need to add fermionic auxiliary fields which split the Yukawa operator in the other possible ways. We introduce Dirac fields $\xi,\xi'$ and a differential operator on such fields $\mathcal{O}^{-1}(-i\partial)$. Then if we add to the Lagrangian
\begin{equation}
\Delta \mathcal{L} = - \overline{\xi} \mathcal{O}^{-1}\xi' - \overline{\xi'} \mathcal{O}^{-1} \xi + c_1 (\overline{\xi} \star \psi \star \varphi + \overline{\psi} \star \xi\star \varphi) + c_2 (\overline{\xi} \star \varphi \star \psi  + \overline{\psi} \star \varphi \star \xi) +  c_3(\overline{\xi'} \psi + \overline{\psi} \xi'),
\end{equation}
we now end up with a contribution to the effective Lagrangian
\begin{equation}
\Delta \mathcal{L}_{\text{eff}} = c_1 c_3 \left( \bar \psi \star \mathcal{O} \left(\psi\right) \star \varphi  + \bar \psi \star \mathcal{O}\left(\psi \star \varphi\right)\right) + c_2 c_3 \left( \bar \psi \star \varphi \star \mathcal{O} \left(\psi\right)   + \bar \psi \star \mathcal{O}\left(\varphi \star \psi \right)\right),
\end{equation}
where we have abused notation and now the argument of $\mathcal{O}$ specifies which fields it acts on. These terms have the right form to correct both vertex orderings.

Now that we've introduced interactions between auxiliary fields and our original fields, the obvious question to ask is whether we can utilize the \textit{same} auxiliary fields to correct both the two-point and three-point actions. In fact, using two auxiliary fields with off-diagonal propagators per particle we may insert any corrections we wish. The new trick is to endow the auxiliary field interactions with extra momentum dependence. 

For a first example with a scalar, consider differential operators $f$, $\Phi$, and add to the Lagrangian
\begin{equation}
\Delta \mathcal{L} = - \chi_1 f^{-1}(-i\partial)\chi_2 + \kappa_1 \chi_1 \varphi + \kappa_2 \chi_2 \overline{\psi} \star \psi + g \varphi\Phi(-i\partial) \chi_2.
\end{equation}
We may now integrate out the auxiliary fields and find 
\begin{equation}
\Delta \mathcal{L}_{\text{eff}} = g \kappa_1 \varphi f(\Phi(\varphi)) + \kappa_1 \kappa_2 \overline{\psi} \star \psi \star f(\varphi) 
\end{equation}
where we've assumed that $f$ and $\Phi$ commute. If we take $\Phi = \mathds{1}$ then we have the interpretation of merely inserting the $\chi$ two-point function in both the two-and three-point functions. But we are also free to use some nontrivial $\Phi$, and thus to make the corrections to the two- and three-point functions have whatever momentum dependence we wish. It should be obvious how to generalize this to insert momentum dependence into the scalar lines of arbitrary $n-$point functions.

The case of a fermion is no more challenging in principle. For differential operators $\mathcal{O}, \mathcal{F}$, we add
\begin{multline}
\Delta \mathcal{L} = - \overline{\xi} \mathcal{O}^{-1}\xi' - \overline{\xi'} \mathcal{O}^{-1} \xi + c_1 (\overline{\xi} \star \psi \star \varphi + \overline{\psi} \star \xi \star \varphi) + c_2 (\overline{\xi} \star \varphi \star \psi  + \overline{\psi} \star \varphi \star \xi ) \\ + c_3(\overline{\xi'} \psi + \overline{\psi} \xi') +  \frac{g}{2} \left( \bar \xi \mathcal{O}^{-1}\mathcal{F}\psi + \bar \psi \mathcal{O}^{-1} \mathcal{F} \xi\right),
\end{multline}
and upon integrating out the auxiliary fields we find
\begin{equation}
\Delta \mathcal{L}_{\text{eff}} = g c_3 \bar \psi \mathcal{F} \psi + c_1 c_3 \left( \bar \psi \star \mathcal{O} \left(\psi\right) \star \varphi  + \bar \psi \star \mathcal{O}\left(\psi \star \varphi\right)\right) + c_2 c_3 \left( \bar \psi \star \varphi \star \mathcal{O} \left(\psi\right)   + \bar \psi \star \mathcal{O}\left(\varphi \star \psi \right)\right),
\end{equation}
where the generalization to $n$-points is again clear. Note that in the fermionic case it's crucial that we be allowed to insert different momentum dependence in the corrections to the two- and three-point functions, as these have different Lorentz structures.

Now we cannot quite implement this for the two- and three-point functions calculated in Section \ref{sec:yukawa}, for the simple reason that we regulated these quantities differently. That is, we have abused notation and the symbol `$\Lambda$' means different things in the results for the two- and three-point functions. In order to carry out this procedure, we could simply regulate the two-point functions in $3d$ Schwinger space, though we run into the technical obstruction that the integration method above only calculates the leading divergence, which is not good enough for the scalar case.